\documentclass[twocolumn]{aastex6}
\usepackage[latin9]{inputenc}
\usepackage{color}
\usepackage{amstext}
\usepackage{graphicx}

\makeatletter

\providecommand{\tabularnewline}{\\}

\usepackage{amsmath}

\newcommand{\be}{\begin{eqnarray}}                                                                                                                                                                                 
\newcommand{\ee}{\end{eqnarray}}

\newcommand{\Chalmers}{Department of Space, Earth and Environment, Chalmers University of Technology, Onsala Space Observatory, 439 92, Onsala, Sweden}
\newcommand{\Curtin}{International Centre for Radio Astronomy Research, Curtin University, Bentley, WA 6102, Australia}

\newcommand{\CSIRO}{CSIRO Astronomy and Space Science, P.O. Box 76, Epping, NSW 1710, Australia}

\makeatother

\begin{document}

\title{Probing pulsar scattering between 120 and 280 MHz with the MWA}

\correspondingauthor{F.~Kirsten}\email{franz.kirsten@chalmers.se}

\author[0000-0001-6664-8668]{F.~Kirsten}\affiliation{\Chalmers}\affiliation{\Curtin}
\author[0000-0002-8383-5059]{N.~D.~R.~Bhat}\affiliation{\Curtin}
\author[0000-0001-8845-1125]{B.~W.~Meyers}\affiliation{\Curtin}  \affiliation{\CSIRO}
\author[0000-0001-6763-8234]{J.-P.~Macquart}\affiliation{\Curtin}
\author[0000-0001-7662-2576]{S.~E.~Tremblay}\affiliation{\Curtin}
\author[0000-0002-6380-1425]{S.~M.~Ord} \affiliation{\CSIRO} 
\begin{abstract}
The high sensitivity and wide frequency coverage of the Murchison
Widefield Array allow for the measurement of the spectral scaling
of the pulsar scattering timescale, $\alpha$, from a single observation.
Here we present three case studies targeted at bright, strongly scattered
pulsars J0534$+$2200 (the Crab pulsar), J0835$-$4510 (the Vela pulsar)
and J0742$-$2822. We measure the scattering spectral indices to be
$-3.8\pm0.2$, $-4.0\pm1.5$, and $-2.5\pm0.6$ for the Crab, Vela,
and J0742$-$2822, respectively. We find that the scattered profiles
of both Vela and J0742$-$2822 are best described by a thin screen
model where the Gum Nebula likely contributes most of the observed
scattering delay. For the Crab pulsar we see characteristically different
pulse shapes compared to higher frequencies, for which none of the
scattering screen models we explore are found to be optimal. The presence
of a finite inner scale to the turbulence can possibly explain some
of the discrepancies. 
\end{abstract}


\section{Introduction}

The interstellar medium (ISM) is a cold plasma that disperses and
scatters pulsar radiation. Dispersion causes the broadband signal
of a pulsar to be smeared out in time such that a particular pulse
arrives at the observer later in time at lower frequencies than at
higher frequencies. This is quantified in terms of the dispersion
measure, ${\rm DM}=\int_{0}^{D}n_{e}(l)\,dl$, which is the integral
over the electron density, $n_{e}$, along the line of sight to a
pulsar at distance $D$. Dispersion is generally well understood,
with the dispersion delay $\tau_{dm}$ scaling as the inverse of the
observing frequency $\nu$ squared: $\tau_{dm}\propto\nu^{-2}$. Thus,
dispersion can be corrected for, modulo the time variability \citep[e.g.][]{keith2013,you2007}
and possible chromaticity \citep{cordes2016} of the ${\rm DM}$.
Both dependencies, ${\rm DM}(t)$ and the suggested ${\rm DM}(\nu)$,
originate in the inhomogeneous, turbulent ISM: the motion of a pulsar
relative to the ISM and Earth causes the pulsar signal to probe different
parts of the ISM over time. At the same time, density variations in
the ISM cause radio waves to be scattered over a range of directions,
resulting in an angular spectrum with a characteristic width $\theta_{{\rm {scatt}}}$,
i.e. the scattering angle. The integrated effect is that an intrinsically
narrow pulse becomes broadened in time as off-axis radiation is scattered
back into the line of sight and arrives later at the observer due
to the extra path length. The measurable effect of scatter broadening,
the scattering delay $\tau$, depends on frequency roughly as $\nu^{-4}$. 

As such, scattering is the main hindrance to perform accurate pulsar
timing at lower frequencies ($\nu\lesssim1\,$GHz) where pulsars tend
to be brighter. At the same time this means that high sensitivity
observations of pulsars at low radio frequencies are ideal to use
both effects and their variation with time and frequency to study
the properties of the ISM itself. The shape of a scattered profile
and its evolution as a function of frequency give us insights into
the underlying geometry, dynamics and physics of the ISM. For example,
the shape of the rising edge of a pulse depends on the distribution
of the scattering medium along the line of sight, namely whether it
is homogeneously distributed or organized in (possibly multiple) thin
sheets. Similarly, the detailed structure of the scattering tail reveals
the underlying structure function of density irregularities in the
ISM \citep{lambert1999}.

\begin{table}
\caption{\label{tab:pulsar-characteristics}Characteristics of the observed
pulsars}

\begin{tabular}{ccccc}
\hline 
 & Period  & DM  & Distance  & (l, b)\tabularnewline
Pulsar & {[}ms{]} & {[}pc$\,$cm$^{-3}${]} & {[}kpc{]} & (\textdegree,\textdegree)\tabularnewline
\hline 
J0534$+$2200  & $33.3$ & $56.7$ & $1.31$\tablenotemark{a} & $(184.5,-5.7)$\tabularnewline
J0742$-$2822  & $166.7$ & $73.7$ & $3.11$\tablenotemark{a} & $(243.7,-2.4)$\tabularnewline
J0835$-$4510 & $89.3$ & $67.9$ & $0.28$\tablenotemark{b} & $(263.5,-2.7)$\tabularnewline
\hline 
\end{tabular}

\tablecomments{From the ATNF Pulsar Catalogue, \citet{Manchester05}}

\tablenotetext{a}{DM-derived distance assuming the YMW16 model of \citet{yao2017}}
\tablenotetext{b}{Parallax based measurement \citep{dodson2003}}
\end{table}

\begin{table*}
\caption{\label{tab:observational-details}Details of the observations}
\begin{tabular}{ccccc}
\hline 
 &  & Observing duration  & Central Frequencies of subbands  & Bandwidth per subband\tabularnewline
Pulsar  & MJD  & {[}seconds{]}  & {[}MHz{]}  & {[}MHz{]} \tabularnewline
\hline 
J0534$+$2200  & 56968  & 3600  & 120.96, 165.76, 210.56, 278.40  & 7.68 \tabularnewline
J0742$-$2822  & 57422  & 300 & 77.44, 149.12, 210.56, 311.68  & 2.56 \tabularnewline
 &  &  & 87.04, 97.28, 107.52, 117.76, 128.00, 158.72  & 1.28 \tabularnewline
 &  &  & 168.96, 179.20, 189.44, 199.68, 220.16,  & 1.28 \tabularnewline
 &  &  & 230.40, 271.36, 281.60, 291.84, 302.08  & 1.28 \tabularnewline
J0835$-$4510 & 56961  & 1200 & 120.96, 210.56, 256.64  & 7.68 \tabularnewline
 &  &  & 164.48  & 5.12 \tabularnewline
 &  &  & 179.84  & 2.56 \tabularnewline
\hline 
\end{tabular}
\end{table*}

Since the impact of the ISM on pulsar radiation is strongest at low
frequencies, observations of bright scattered pulsars at $\nu\lesssim300\,$MHz
are well suited to differentiating between competing models of the
ISM. In particular, it remains unclear whether Kolmogorov turbulence
appropriately describes and quantifies variations in ${\rm DM}$,
the frequency scaling of pulsar scattering, and the scintillation
properties of pulsars \citep[e.g.][]{lam2016}. The parabolic arcs
in secondary spectra of scintillating pulsars \citep[e.g.][]{brisken2010,stinebring2001}
disfavor a homogeneous, isotropically distributed scattering medium
\citep[e.g.][]{walker2004}. Instead, they hint at localized, anisotropic
scattering structures that could have a sheet like geometry \citep[e.g.][]{braithwaite2015,pen2014}.

The scattering delay induced by the interstellar medium scales with
frequency as $\nu^{-2\beta/(\beta-2)}$, where $\beta$ is the power
spectral index of density irregularities. For Kolmogorov turbulence
we expect $\beta=11/3$ \citep[e.g.][]{rickett1990}. Most of the
published measurements of $\alpha=-2\beta/(\beta-2)$, however, deviate
from the theoretically expected value $\alpha=-4.4$. In fact, the
majority of recently published measurements obtained from low frequency
observations of pulsars list spectral indices $\alpha>-4$ \citep{geyer2017,meyers2017,krishnakumar2017,eftekhari2016,lewandowski2015}.
Evidence for shallower scaling also comes from a large sample of pulsar
measurements by \citet{bhat2004}, who deduced a value of $\langle\alpha\rangle=-3.9\pm0.2$
for the global scattering index.\textcolor{red}{{} }They also offer
a possible explanation in terms of an inner scale and a crossover
point that depends on this scale. \citet{cordes2001} explain such
`anomalous' scattering behavior by introducing filamentary structures
along the line of sight; a scenario that can also give rise to shallow
scattering. \citet{xu2017} invoke supersonic turbulence and a frequency
dependent volume filling factor of density irregularities to reconcile
observations and theoretical models. \citet{2016MNRASGeyer} investigate
the idea of anisotropic pulse broadening functions in the case of
a thin screen, finding that the standard isotropic models can significantly
underestimate the scattering time, especially at low frequencies.
In a recent study, \citet{geyer2017} compare spectral indices obtained
by fitting isotropic and anisotropic models to multi-frequency scattering
observations, finding that the latter models yield values for $\alpha$
that are in better agreement with theoretical expectations. 

Obviously, the ISM cannot be fully described by an isotropic model
in which the power spectrum of density irregularities follows a simple
power law. In order to shed further light on the physics of the ISM,
more detailed studies of the shape and the temporal and spectral evolution
of scattered pulsar profiles at low frequencies ($\nu\lesssim300\,$MHz)
are required. New instruments such as the Murchison Widefield Array
\citep[MWA,][]{tingay13}, the Low Frequency Array \citep[LOFAR,][]{vanHaarlem2013},
and the Long Wavelength Array \citep[LWA,][]{tayor2012} can provide
the sensitivity, the frequency coverage, and also the temporal resolution
needed for making further progress in this area.

In this work we present a pilot study of pulsar scattering using the
MWA. We selected three bright, moderate DM pulsars \textendash{} J0534$+$2200,
J0742$-$2822, and J0835$-$4510 (Table \ref{tab:pulsar-characteristics})
\textendash{} and observed them across the large frequency range of
the MWA ($80-300\,$MHz). We show that for bright pulsars the sensitivity
of the MWA is sufficient to obtain high quality pulse profiles within
a short integration time ($\sim5\,$min), allowing for a measurement
of the frequency scaling index $\alpha$ for all three targets from
a single observation. 

In Section 2 we describe the MWA observations, and in Section 3 we
present our results and an analysis of the scattering profiles. Our
results are discussed in Section 4 and the conclusions are presented
in Section 5.

\section{Observations \& data reduction}

\begin{figure*}
\centering{}\includegraphics[width=0.49\textwidth]{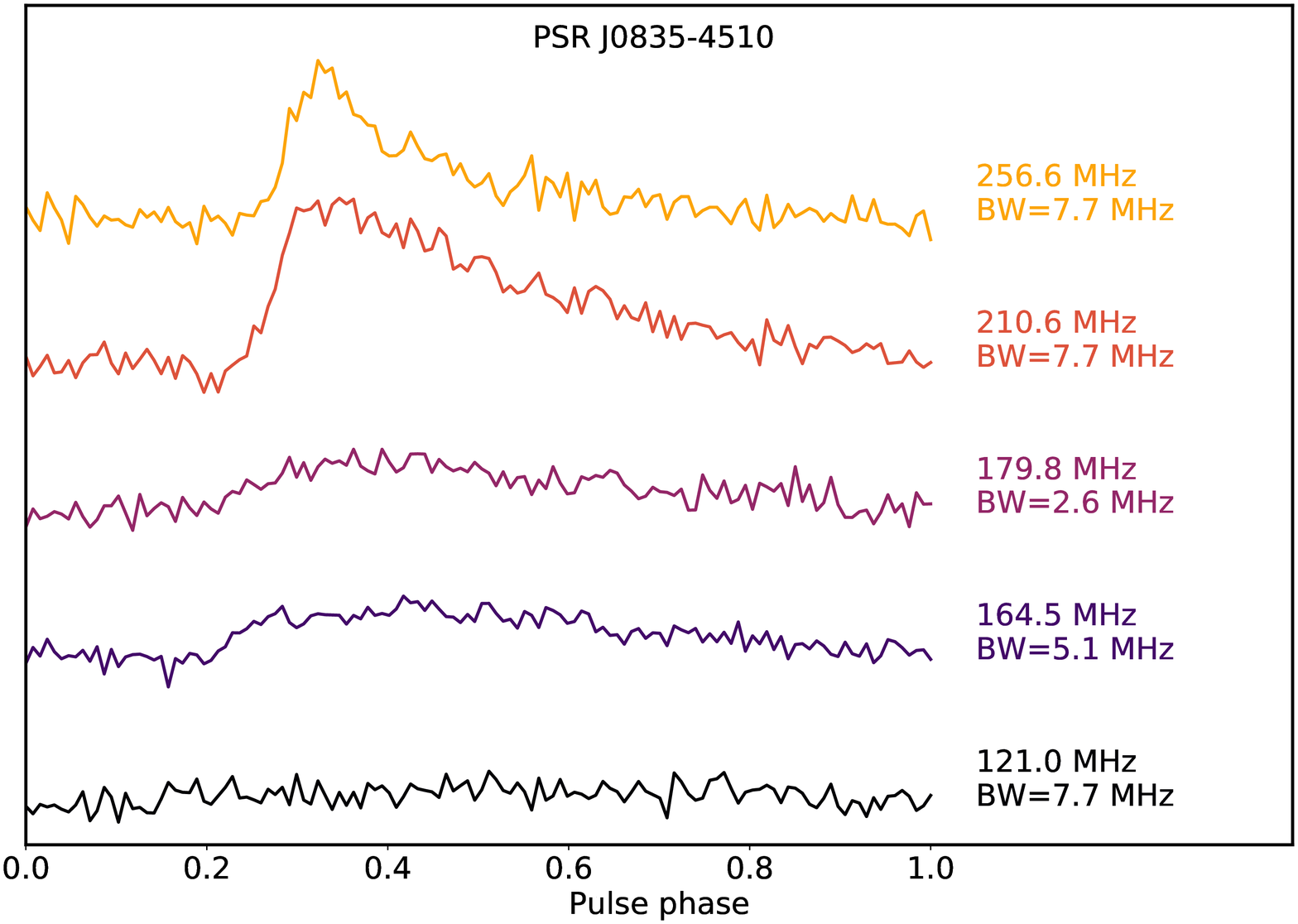}
\includegraphics[width=0.49\textwidth]{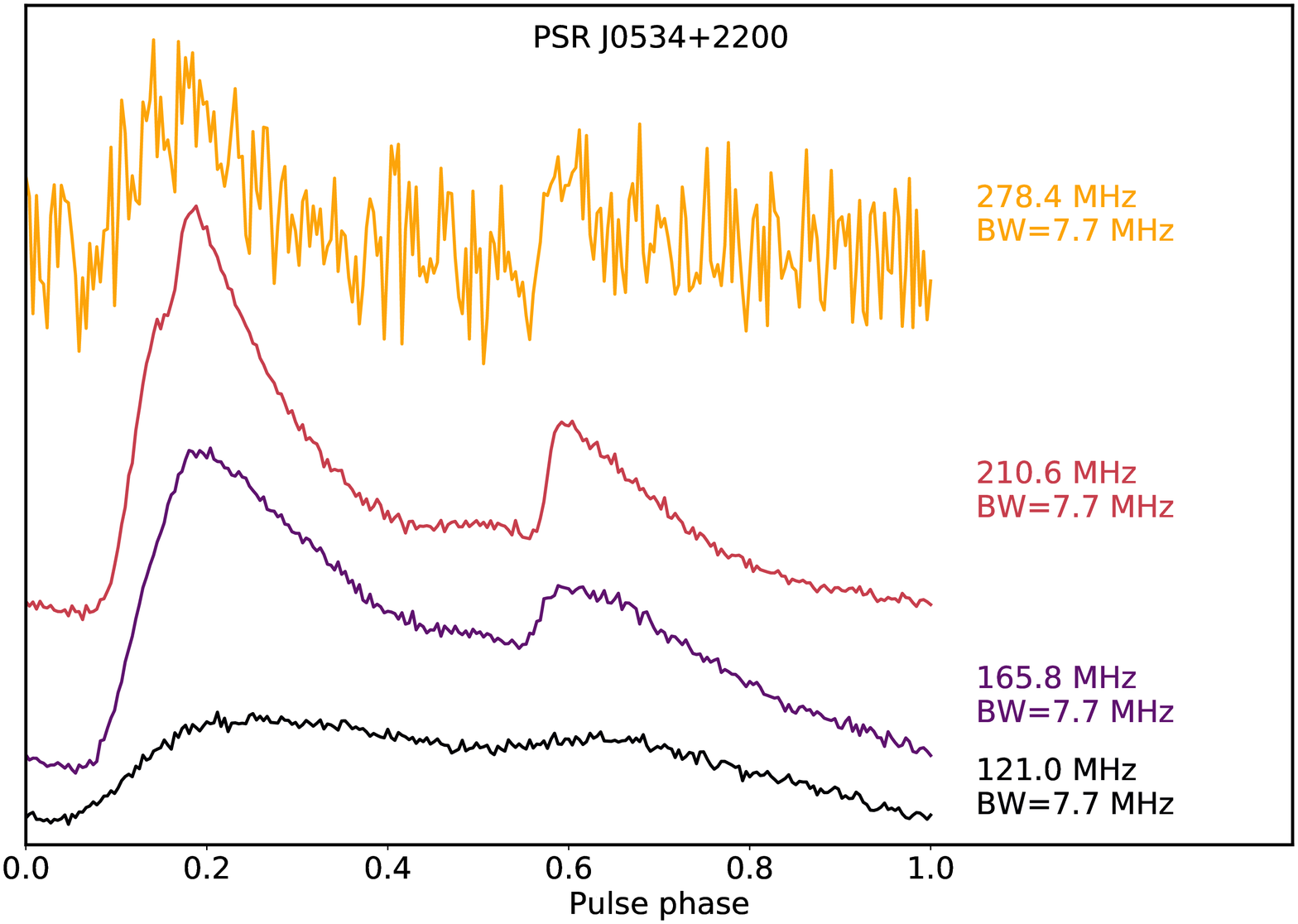} \caption{\label{fig:vela+crab-stacked-freq}Folded pulse profiles of the Vela
pulsar (left) and the Crab pulsar (right) across the MWA-band. We
used coherently summed data for frequencies below $235\;$MHz while
we used incoherently summed data for the highest bands (we did not
manage to form a coherent beam above $235\;$MHz because of RFI).
For Vela, the bandwidth varies as indicated while for the Crab it
is constant at $7.7\;$MHz. The low signal-to-noise ratio (SNR) for
the Crab at $278.4\;$MHz is due to three factors: the steep spectrum
of the pulsar, the amount of RFI in the band, and the reduced sensitivity
of the MWA towards the upper end of the band.}
\end{figure*}

We observed the three pulsars J0534$+$2200 (the Crab pulsar), J0742$-$2822,
and J0835$-$4510 (the Vela pulsar) with the MWA using the Voltage
Capture System (VCS, \citealt{tremblay15}). This observing mode records
the channelised voltages from each of the 128 tiles\footnote{The tiles are about $4\,\mathrm{m}\times4\,\mathrm{m}$ in size, comprising
16 dipoles each. At the time of the observations, the majority of
the tiles (112) were spread over a $1.5\,$km core region with the
remaining 16 tiles being located at larger distances, allowing for
baselines up $3\,$km in length.}, which can be summed coherently after performing calibration. In
order to be able to characterize the scattering properties of each
target we observed in the following manner: instead of observing over
a contiguous band of maximal $30.72\,$MHz bandwidth, we spread the
24 available individual coarse channels across the accessible frequency
range of $80-300\,$MHz.\textbf{ }Each of these coarse channels has
a bandwidth of $1.28\,$MHz which, before recording, is channelised
into 128 fine channels (i.e. $10\,$kHz) in a polyphase filterbank.
Multiple subbands can be constructed by either grouping consecutive
coarse channels, or by using individual coarse channels as subbands.

We used different frequency setups for each target as indicated in
Table \ref{tab:observational-details}. The number of subbands ranged
from 4 to 20, hence the bandwidth varied between $1.28\,$MHz and
$7.68\,$MHz per subband. Observing duration varied from about 5 to
60 minutes depending on the target. For calibration purposes we also
performed a short ($\sim2\,$min) observation of a strong nearby calibrator
(non-pulsar continuum source) using the same frequency setup as for
the respective pulsar prior to the target observations. The calibrator
scans were not recorded with the VCS but instead with the MWA correlator
\citep{ord15} which produced visibilities at a cadence of $10\,$kHz
and $2\,$s.

\subsection{Calibration}

We performed calibration using the Real Time System (RTS, \citealt{mitchell08})
which corrects for various instrumental effects and removes positional
offsets of sources induced by the refractive ionosphere. The RTS uses
a fully polarimetric calibration formalism \citep{sault1996} to form
a least squares estimate of the complex gain of the constituent antennas
in the array. Ionospheric refraction manifests predominantly as a
position offset in the direction of the calibrator, and can be extracted
from the antenna phase terms as a phase ramp with a quadratic frequency
dependence. Thus antenna gain terms are obtained that are direction
independent and can be applied to form phased array beams anywhere
on the sky.

The RTS generates bandpass calibration solutions and direction-independent
Jones Matrices (DIJs). The latter contain full polarization calibration
solutions across the entire field of view (FOV) of the MWA (Full Width
Half Maximum$\sim25^{\circ}$ at $150\,$MHz) for each of the individual
24 channels. The RTS inherently assumes a contiguous bandpass of $30.72\,$MHz
bandwidth which is not directly applicable in the case of split band
observations performed here. It is, however, possible to calibrate
individual channels or groups of consecutive channels. We thus used
the RTS software package separately for each of the subbands listed
in Table \ref{tab:observational-details} using models of the respective
calibrators as developed by \citet{line16}. Typically, the ionosphere
is stable enough for one such calibration solution to be valid over
several hours.

In the case of the Crab pulsar we did not use the dedicated calibrator
observation but instead performed in-field calibration on the target
itself. As the RTS performs its analysis on visibilities, we first
correlated 200 seconds of the voltages recorded for the Crab pulsar
using the standard MWA software correlator.

\subsection{Coherent beam-forming}

\begin{figure*}
\includegraphics[width=1\textwidth]{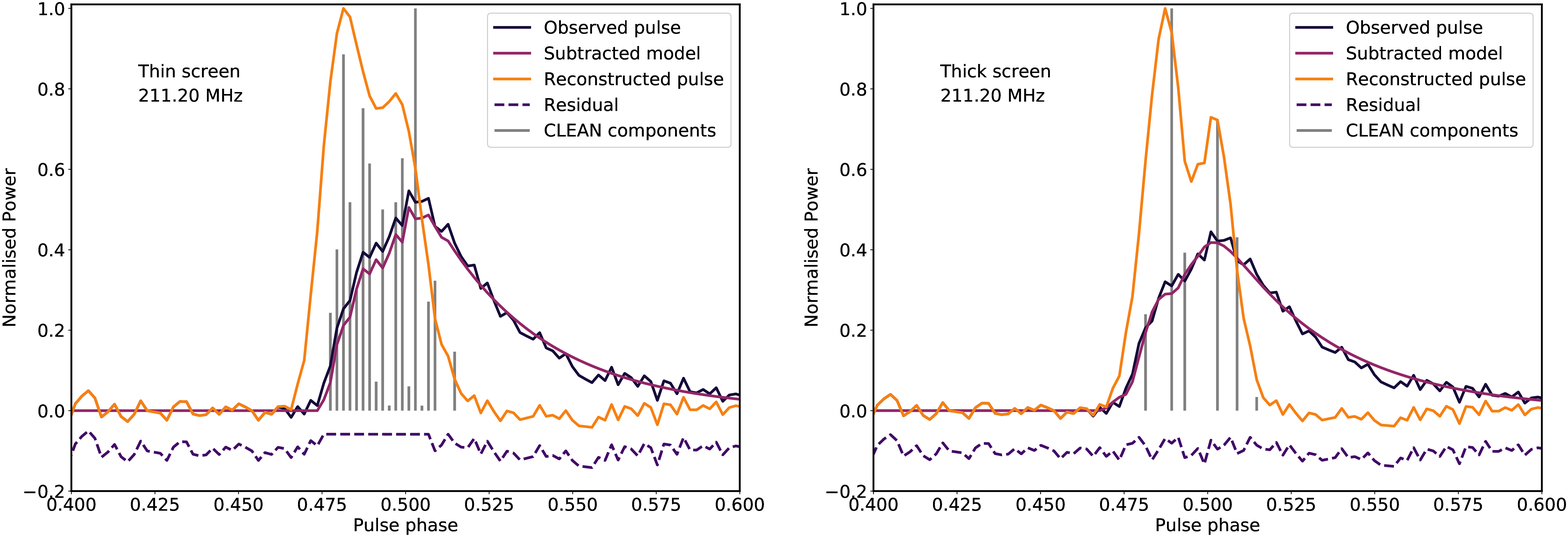}

\caption{\label{fig:clean-procedure}Illustration of the CLEANing process using
the example of J0742$-$2822 at $211.2\,$MHz for both the thin (left)
and thick (right) screen models. The black line shows the observed
pulse, the vertical gray lines represent the locations and relative
amplitudes of the CLEAN components summed up per phase bin. The purple
solid line shows the sum of all CLEAN components that were convolved
with the thin/thick screen model and subsequently subtracted from
the observed pulse. The residual after subtraction is represented
by the purple dashed line and is offset from the origin for illustration
purposes. Finally, the yellow line shows the recovered (intrinsic)
pulse shape as computed from the CLEAN components convolved with a
Gaussian.}
\end{figure*}

To compute the coherent sum of the signals from all 128 tiles, we
use a software package specifically developed for the MWA (Ord et
al., in prep., referred to as the `beamformer' below). The full description
of the beamformer is beyond the scope of this article but we describe
the individual steps here briefly.

When forming the coherent sum of the signals from all tiles the FOV
of the MWA is reduced significantly. Between the lower ($\sim80\,$MHz)
and the upper ($\sim300\,$MHz) band edges of the instrument, the
size of this pencil beam ranges from about $4\arcmin$ to about $1\arcmin$,
which is, effectively, the size of the synthesized beam. To phase
up on a particular field of this size within the incoherent FOV, the
pencil beam needs to be `steered' towards that desired direction.
As the name suggests, the DIJs obtained after running the RTS on the
calibrator scan contain no information about a specific direction
(i.e. towards a source) within the incoherent FOV of the target scan.
Therefore, starting from the DIJs we first compute the geometric phase
offsets from the pointing center of the target scans for the direction
we would like to form a coherent beam on (i.e. the coordinates of
the pulsar). These delays are computed for each $10\,$kHz wide fine
channel on a per-second basis. The complex gains from both the DIJs
and the direction-dependent calibration are applied to the raw voltages
of the target scans, after which the signals are summed and detected
to form a time series of power.

This strategy works provided the RTS converges on a valid calibration
solution. In our observations, frequency subbands above $\sim235\,$MHz
were, unfortunately, strongly affected by radio frequency interference
(RFI) generated by satellites. As a result, the RTS failed to provide
usable DIJs, thus making a coherent summation of the raw voltages
impossible. Therefore we reverted to the incoherently summed voltages\footnote{For the incoherent sum we detect the signal per tile and then sum
the time series of power. This reduces the sensitivity (compared to
the coherent beam) by a factor $\sqrt{{N}}$, where $N$ is the number
of tiles.} in bands above $235\,$MHz. As the Crab and Vela were sufficiently
bright and integrations times were long, we were able to detect both
pulsars even with the MWA's incoherent sensitivity. For J0742$-$2822
the integration time and bandwidth were too small to result in a detection
in this upper frequency range using the incoherently summed data.

\section{Analysis \& Results\label{sec:Analysis-=000026-Results}}

\begin{figure*}
\includegraphics[height=0.33\textwidth]{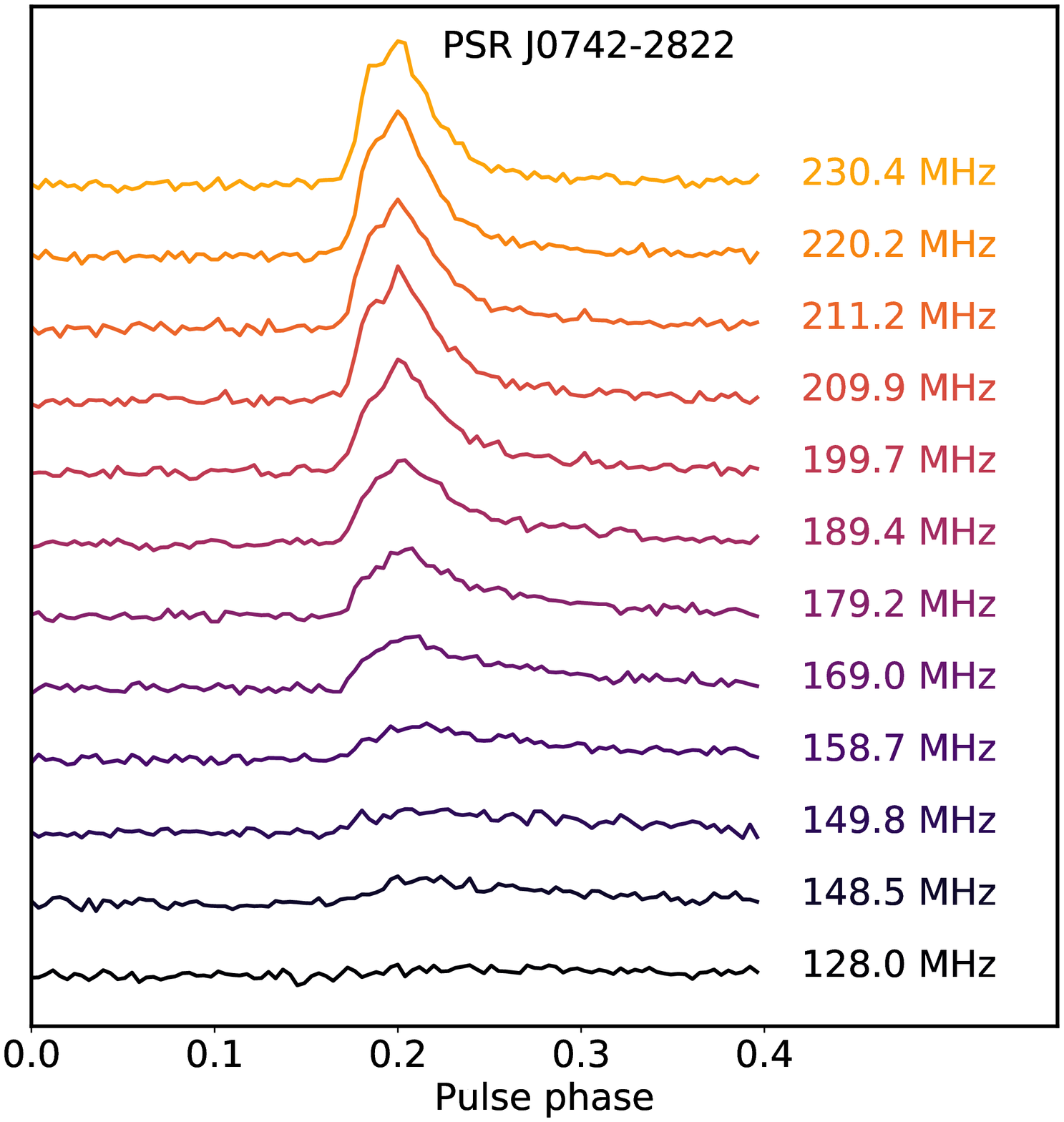} \includegraphics[height=0.33\textwidth]{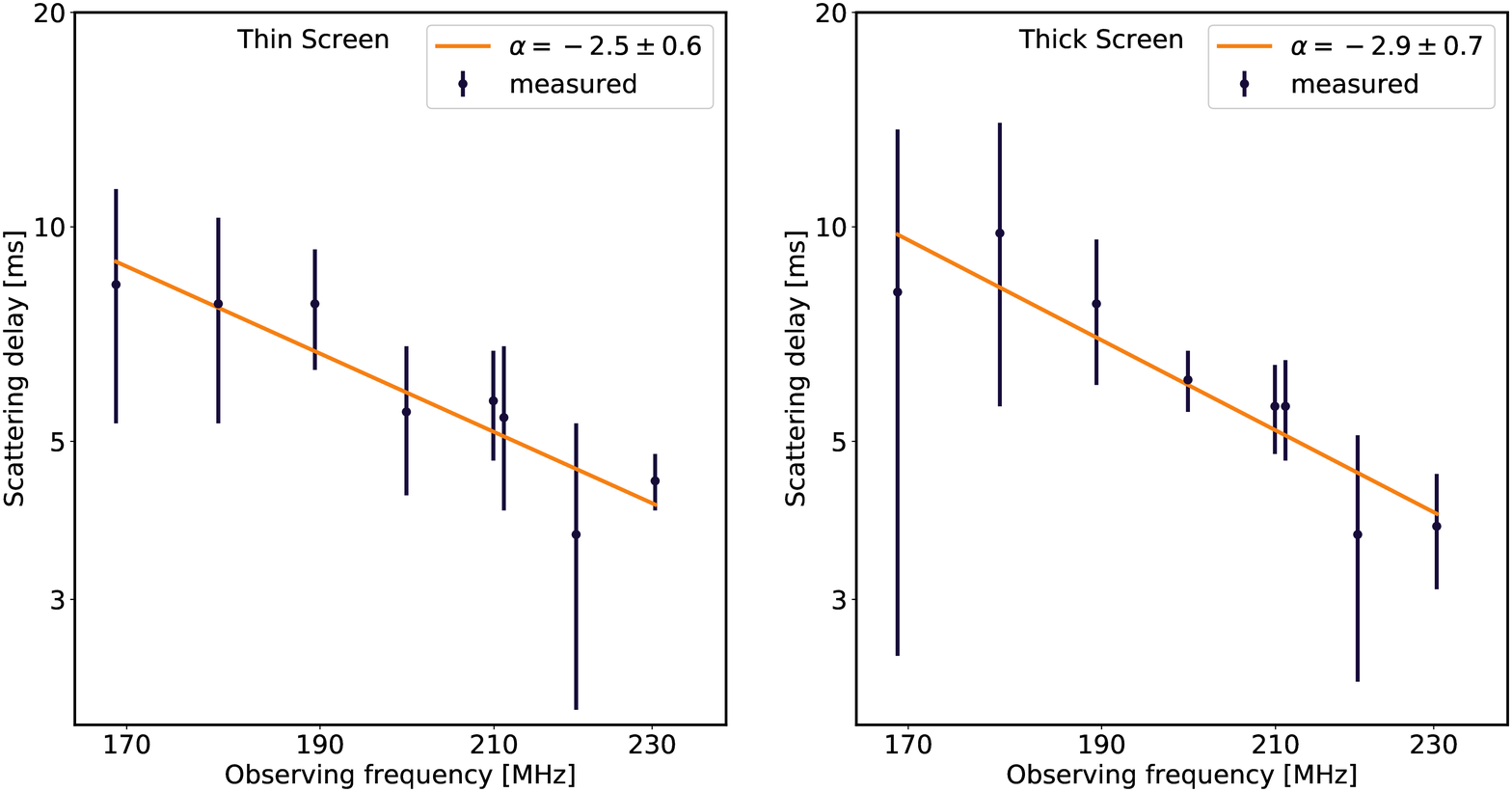}
\caption{\label{fig:0742-stacked-freq}Left: Measured pulse profiles of J0742$-$2822
as seen in coherently summed data, zoomed in on the pulse phase containing
the signal. The bandwidth in each subband is $1.28\,$MHz. Below an
observing frequency of $148.5\,$MHz the pulsar is scattered beyond
detection while above $230\,$MHz RFI prevented us from coherently
summing the array. Middle and right: Measured scattering delay vs.
observing frequency for J0742$-$2822 using both a thin and a thick
screen model for scattering, as described in the text. The yellow
line in both plots is a power law fit of the form $\tau\propto\nu^{\alpha}$.
In the fitting procedure we omitted the measurements below $169\,$MHz
because of low SNR and, hence, unreliable scattering delays.}
 
\end{figure*}

We used the software packages PSRCHIVE \citep{hotan04,vanStraten2012}
and DSPSR \citep{vanStraten2011} for further data processing and
analysis. We processed the individual subbands for each pulsar separately,
and the resultant, incoherently dedispersed pulse profiles are shown
in Figures \ref{fig:vela+crab-stacked-freq} and \ref{fig:0742-stacked-freq}. 

One of our main goals is to quantify the physical nature of scattering
along the sight lines to the three different pulsars. To that end,
we explore different pulse broadening functions (PBFs) to fit the
observed pulse profiles in each band and measure the frequency scaling
of the scattering delay $\tau$. However, for a least-squares fitting
procedure to be easily applicable, the intrinsic pulse shape needs
to be known with high accuracy and the characteristic scattering time
needs to be small enough that the scattering tail does not extend
beyond the pulsar period. In practice, this need not necessarily be
the case; measured profiles can be composed of multiple components
whose shapes and separations cannot be easily extrapolated to low
frequencies from higher frequency observations. Moreover, long scattering
tails result in an apparent merging of the components and render the
determination of a baseline impossible. In the case of the Crab pulsar
this is alleviated by analyzing individual giant pulses \citep[e.g. ][]{meyers2017,bhat2007}. 

For the other two pulsars we employ a deconvolution technique based
on the CLEAN algorithm \citep{hogbom74} which was implemented for
pulse profiles by \citet{bhat03}. The method recovers an intrinsic
pulse shape by deconvolving the observed profile with an assumed model
of the PBF (Fig. \ref{fig:clean-procedure}), nonetheless trialling
over a range of $\tau$ to determine the best-fit value. Two commonly
adopted models are a thin screen model and a thick screen model.

In general, the exact form of the PBF associated with the ISM is unknown
and depends on the distribution and the turbulence power spectrum
of the scattering material along the line of sight to the pulsar.
Assuming a square-law structure function for density inhomogeneities,
the simplest PBF for a thin screen model is a one-sided exponential,
\begin{equation}
g_{0}(t)=\left(\frac{1}{\tau}\right)\exp\left[-\frac{t}{\tau}\right]u(t),\label{eq:thinscreen}
\end{equation}
where $u(t<0)=0$ and $u(t\geq0)=1$. 

More realistic models typically have a shallower rise time and a longer
exponential tail, corresponding to different peak positions. \citet{1972MNRASWilliamson}
showed that for a thick scattering screen, the PBF at small times
$t$, i.e. during the rise time up until shortly after the peak of
the emission, is given by
\begin{equation}
g_{1}(t)=\left(\frac{\pi\tau}{4t^{3}}\right)^{1/2}\exp\left[-\frac{\pi^{2}\tau}{16t}\right].\label{eq:thickfinite}
\end{equation}
The shape of the decaying tail can be described by an exponential,
$e^{-t/\tau}$, similar to $g_{0}(t)$. 

To evaluate the quality of the deconvolution procedure we make use
of the figures of merit (FOM) defined in \citet{bhat03}. In particular
these are the positivity $f_{r}$ (the PBF should not over-subtract
the flux), the symmetry $\Gamma$ (the deconvolved pulse should not
show a residual scattering tail) and the consistency of the residuals
with the off-pulse rms indicated by $N_{\mathrm{r}}$ (the relative
number of bins that are within three standard deviations of the off-pulse
region) and $\sigma_{\mathrm{r}}$ (the ratio of the rms in the CLEANed
window to the rms in the off-pulse region). For convenience we summarize
the definitions of each parameter in Appendix \ref{app:Figures-of-Merit}.
The deconvolution procedure should minimize $f_{r}$ and $\Gamma$,
while maximizing $N_{\mathrm{r}}$ and achieving $\sigma_{\mathrm{r}}\sim1$.
It is important to note that the absolute numbers of parameters $\Gamma$,
$N_{\mathrm{r}}$ and $\sigma_{\mathrm{r}}$ do not allow for a meaningful
assessment which scattering model provides a better fit. They merely
serve to find the best fit $\tau$ as illustrated in Figure \ref{fig:Examples-of-the-fom}.

We use $f_{r}$ also to estimate the uncertainties of our measured
$\tau$. In a least-squares fitting routine, the minimum of the reduced
chi-square statistic, $\chi_{\mathrm{red}}^{2}$, indicates the best
fit model. The parameter range for which $\chi_{\mathrm{red}}^{2}$
is unity above that minimum indicates the uncertainty of the measured
parameter. Similarly, we use the range in $\tau$ for which $f_{r}$
is unity above the best fit model to estimate our uncertainties. In
general the absolute number of this parameter carries little information.
It is only in comparison to another model where a lower value of $f_{r}$
indicates a better fit. Similary, we also compute a reduced chi-square
value, $\chi_{\mathrm{nCC}}^{2}$, from our best-fit model in order
to have a relative measure for the goodness of fit between the thick
screen model and the thin screen model. In this context we use the
number of pulse phase bins in which CLEAN components were subtracted,
$nCC$, as an estimate for the number of degrees of freedom\footnote{The total number of CLEAN components is a strong function of the loop
gain, the S/N of the pulse, and the time resolution of the data; it
is typically of the order of a few thousand.}. We emphasize that $\chi_{\mathrm{nCC}}^{2}$ computed in this way
is merely another figure of merit that can be employed to prefer one
model for scattering over another.

\subsection{PSR J0742$-$2822}

\begin{table*}
\caption{\label{tab:0742-scattering-times} Measured scattering delays for
J0742$-$2822}
\begin{tabular}{cccccccc|ccccccc}
\hline 
Frequency & \multicolumn{7}{c}{Thin screen} & \multicolumn{7}{c}{Thick screen}\tabularnewline
{[}MHz{]}  & $\tau\;${[}ms{]}  & $f_{r}$  & $N_{\mathrm{r}}$ & $\mathbf{\sigma_{\mathrm{r}}}$ & $\Gamma$ & $\chi_{\mathrm{nCC}}^{2}$ & $nCC$ & $\tau\;${[}ms{]}  & $f_{r}$ & $N_{\mathrm{r}}$ & $\mathbf{\sigma_{\mathrm{r}}}$ & $\Gamma$ & $\chi_{\mathrm{nCC}}^{2}$ & $nCC$\tabularnewline
\hline 
$230.40$ & $4.4(4)$  & $0.77$ & $0.19$ & $1.95$ & $0.22$ & $42$ & $17$ & $3.8(7)$  & $0.55$ & $0.19$ & $1.25$ & $0.31$ & $112$ & $6$\tabularnewline
$220.16$ & $3.7(16)$  & $0.12$ & $0.18$ & $1.23$ & $0.55$ & $29$ & $21$ & $3.7(14)$  & $0.63$ & $0.18$ & $1.13$ & $0.54$ & $89$ & $7$\tabularnewline
$211.20$ & $5.4(14)$  & $0.12$ & $0.18$ & $0.89$ & $0.31$ & $32$ & $18$ & $5.6(9)$  & $0.20$ & $0.19$ & $0.89$ & $0.19$ & $91$ & $6$\tabularnewline
$209.92$ & $5.7(9)$  & $0.30$ & $0.19$ & $1.23$ & $0.23$ & $32$ & $17$ & $5.6(8)$  & $0.62$ & $0.19$ & $0.97$ & $0.08$ & $110$ & $5$\tabularnewline
$199.68$ & $5.5(13)$  & $0.11$ & $0.19$ & $1.57$ & $0.02$ & $45$ & $18$ & $6.1(6)$  & $0.70$ & $0.19$ & $1.13$ & $-0.19$ & $129$ & $6$\tabularnewline
$189.44$ & $7.8(15)$  & $0.05$ & $0.20$ & $2.03$ & $0.29$ & $40$ & $14$ & $7.8(18)$  & $0.14$ & $0.19$ & $1.00$ & $0.29$ & $172$ & $3$\tabularnewline
$179.20$ & $7.8(25)$  & $0.02$ & $0.19$ & $0.90$ & $0.50$ & $36$ & $14$ & $9.8(42)$  & $0.37$ & $0.20$ & $0.91$ & $-0.29$ & $147$ & $3$\tabularnewline
$168.96$ & $8.3(30)$  & $0.03$ & $0.19$ & $1.16$ & $0.22$ & $46$ & $15$ & $8.1(56)$  & $0.05$ & $0.19$ & $1.03$ & $0.63$ & $129$ & $5$\tabularnewline
\hline 
$\alpha$  & \multicolumn{7}{c}{$-2.5(6)$} & \multicolumn{7}{c}{$-2.9(7)$}\tabularnewline
\hline 
\end{tabular}

\tablecomments{Numbers in brackets indicate the uncertainty in the last digit. }
\end{table*}

\begin{table*}
\caption{\label{tab:vela-scattering-times} Measured scattering delays for
the Vela Pulsar}
\begin{tabular}{cccccccc|ccccccc}
\hline 
Frequency & \multicolumn{7}{c}{Thin screen} & \multicolumn{7}{c}{Thick screen}\tabularnewline
{[}MHz{]}  & $\tau\;${[}ms{]}  & $f_{r}$  & $N_{\mathrm{r}}$ & $\mathbf{\sigma_{\mathrm{r}}}$ & $\Gamma$ & $\chi_{\mathrm{nCC}}^{2}$ & $nCC$ & $\tau\;${[}ms{]}  & $f_{r}$ & $N_{\mathrm{r}}$ & $\mathbf{\sigma_{\mathrm{r}}}$ & $\Gamma$ & $\chi_{\mathrm{nCC}}^{2}$ & $nCC$\tabularnewline
\hline 
$256$  & $14.1\pm1.7$  & $1.6$  & $0.26$ & $1.66$ & $0.33$ & $21$ & $5$ & $11.4\pm2.5$  & $0.53$ & $0.28$ & $1.65$ & $0.36$ & $21$ & $5$\tabularnewline
$210$  & $31.2\pm8.1$  & $0.2$  & $0.31$ & $1.24$ & $0.86$ & $22$ & $5$ & $32.6\pm4.7$  & $0.28$ & $0.31$ & $1.46$ & $1.13$ & $38$ & $3$\tabularnewline
\hline 
$\alpha$  & \multicolumn{7}{c}{$-4.0\pm1.5$} & \multicolumn{7}{c}{$-5.0\pm1.5$}\tabularnewline
\hline 
\end{tabular}
\end{table*}

Using the above method, we analyzed the J0742$-$2822 data by adopting
a PBF form as in Eq. \ref{eq:thinscreen} and Eq. \ref{eq:thickfinite}.
We have performed this for each subband separately; the resultant
scattering delays $\tau$ and associated FOM are summarized in Table
\ref{tab:0742-scattering-times}. Overall, the positivity parameter
$f_{r}$ and the parameter $\chi_{\mathrm{nCC}}^{2}$ are lower for
the thin screen model than the thick screen one indicating that a
thin screen geometry is a better representation of the data. 

The middle and right panels of Figure \ref{fig:0742-stacked-freq}
depict the measured scattering delays as a function of frequency for
both the thin and thick screen models, respectively. Also shown are
power law fits of the form $\tau\propto\nu^{\alpha}$. For the thin
screen model we obtain $\alpha=-2.5\pm0.6$, which is well in agreement
with the measurements of \citet[$\alpha=-2.52\pm0.3$]{lewandowski2015}
but lower than those of \citet{geyer2017} who measure $\alpha=-3.8\pm0.4$.
The use of the thick screen model, yields a slightly higher $\alpha=-2.9\pm0.7$,
which agrees with the thin-screen-estimate within the uncertainties.

\begin{figure*}
\includegraphics[width=0.495\textwidth]{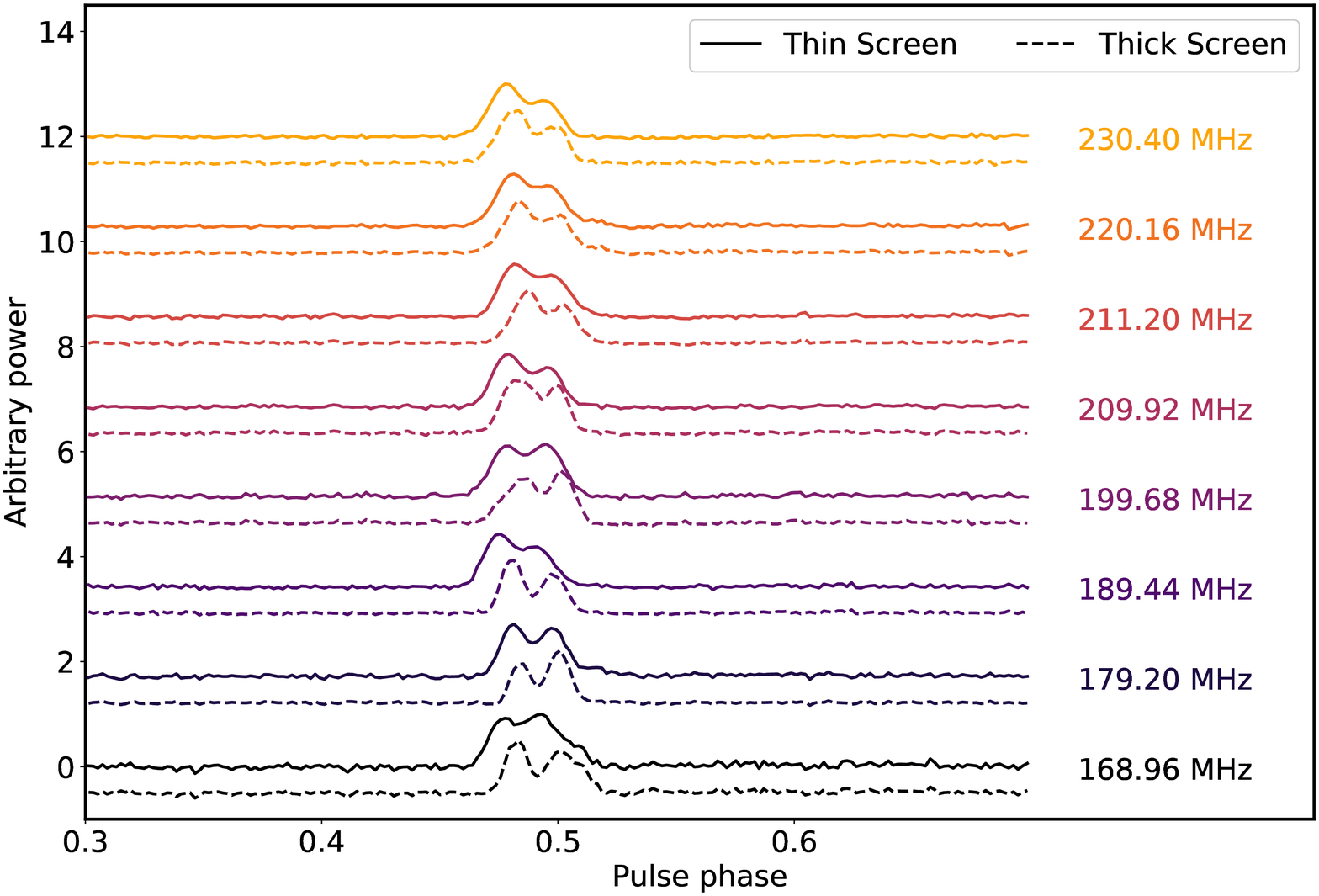}\includegraphics[width=0.495\textwidth]{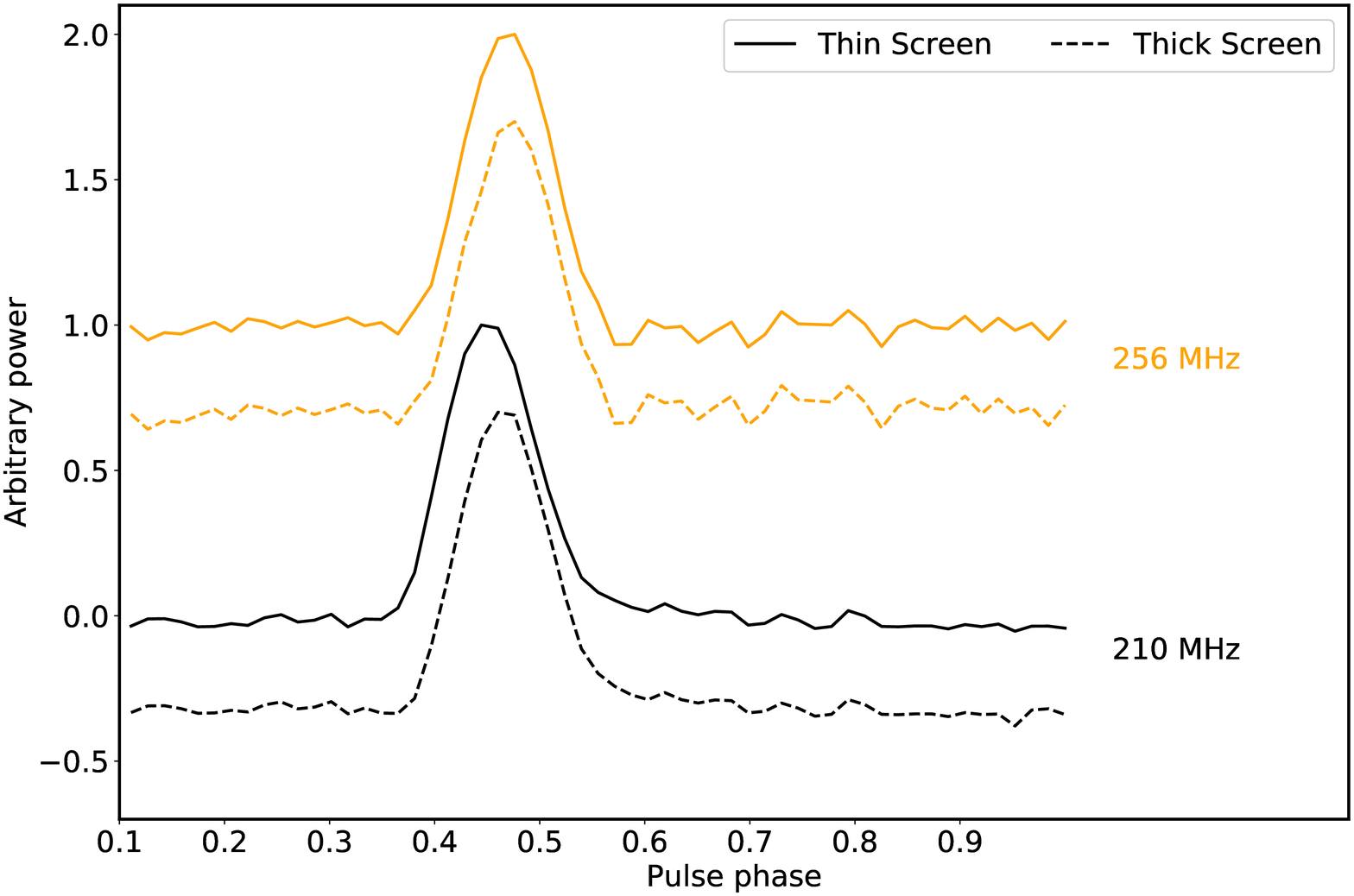}

\caption{\label{fig:reconstructed pulses}Reconstructed pulse profiles for
PSR J0742$-$2822 (left) and the Vela pulsar (right). The solid lines
show the results from using the thin screen model while the dashed
lines show those from the thick screen model. The different observing
bands are color coded.}
\end{figure*}

\subsection{The Vela Pulsar}

In the lowest band of our observations, the effect of scatter broadening
becomes too large to enable a detection of the pulsar, and the detection
in the two bands centered at $164.5\,$MHz and $179.8\,$MHz is only
marginal (Figure \ref{fig:vela+crab-stacked-freq}). Therefore, we
restrict our analysis to the highest two bands, centered on $210.56\,$MHz
and $256.64\,$MHz. The frequency scalings as indicated by the measured
delays (Table \ref{tab:vela-scattering-times}) are $\alpha=-4.0\pm1.5$
(thin screen model) and $\alpha=-5.0\pm1.5$ (thick screen model).
The FOM also listed in Table \ref{tab:vela-scattering-times} do not
indicate a clear trend for which model reproduces the observations
better. The post-CLEANing residuals are similar in their noise characteristics
(indicated by $\sigma_{\mathrm{r}}$ and $N_{\mathrm{r}}$) and the
skewness $\Gamma$ is also very similar between the models. Both the
$f_{r}-$parameter and $\chi_{\mathrm{nCC}}^{2}$ indicate that the
thick screen model is a better representation of the observations
at $256\,$MHz, while at $210\,$MHz the thin screen model provides
a better fit. 

\subsection{The Crab Pulsar\label{subsec:The-Crab-Pulsar}}

\begin{table*}
\caption{\label{tab:crab-fitted-delays}Measured scattering delays for the
Crab Pulsar and implied frequency scalings}

\begin{tabular}{cccccccccccccc}
\hline 
 & \multicolumn{6}{c}{Square law structure function} &  & \multicolumn{6}{c}{Fully diffractive Kolmogorov turbulence}\tabularnewline
\hline 
Frequency & \multicolumn{2}{c}{Thin} & \multicolumn{2}{c}{Thick} & \multicolumn{2}{c}{Modified thin} &  & \multicolumn{2}{c}{Thin} & \multicolumn{2}{c}{Thick} & \multicolumn{2}{c}{Double thin}\tabularnewline
{[}MHz{]}  & $\tau\;${[}ms{]}  & $\chi_{\mathrm{red}}^{2}$  & $\tau\;${[}ms{]}  & $\chi_{\mathrm{red}}^{2}$  & $\tau\;${[}ms{]}  & $\chi_{\mathrm{red}}^{2}$  &  & $\tau\;${[}ms{]}  & $\chi_{\mathrm{red}}^{2}$  & $\tau\;${[}ms{]}  & $\chi_{\mathrm{red}}^{2}$  & $\tau\;${[}ms{]}  & $\chi_{\mathrm{red}}^{2}$\tabularnewline
\hline 
121  & $58(1)$  & $0.97$  & $44.9(9)$  & $0.85$  & $40(2)$  & $0.95$  &  & $47(13)$  & $1.26$  & $46(2)$  & $1.16$  & $47(13)$  & $1.26$\tabularnewline
165  & $18.3(2)$  & $1.06$  & $12.8(1)$  & $1.08$  & $11.9(2)$  & $1.01$  &  & $14(2)$  & $1.44$  & $13.2(3)$  & $1.47$  & $14(2)$  & $1.52$\tabularnewline
210  & $8.4(1)$  & $1.16$  & $5.72(6)$  & $1.01$  & $4.66(9)$  & $1.08$  &  & $6.4(9)$  & $1.58$  & $6.75(7)$  & $1.40$  & $6.3(9)$  & $1.84$\tabularnewline
\hline 
$\alpha$  & \multicolumn{2}{c}{$-3.5(1)$} & \multicolumn{2}{c}{$-3.8(2)$} & \multicolumn{2}{c}{$-3.9(1)$} &  & \multicolumn{2}{c}{$-3.7(2)$} & \multicolumn{2}{c}{$-3.6(3)$} & \multicolumn{2}{c}{$-3.7(2)$}\tabularnewline
\hline 
\end{tabular}
\end{table*}

\begin{figure*}[!t]
\begin{centering}
\includegraphics[width=0.5\textwidth]{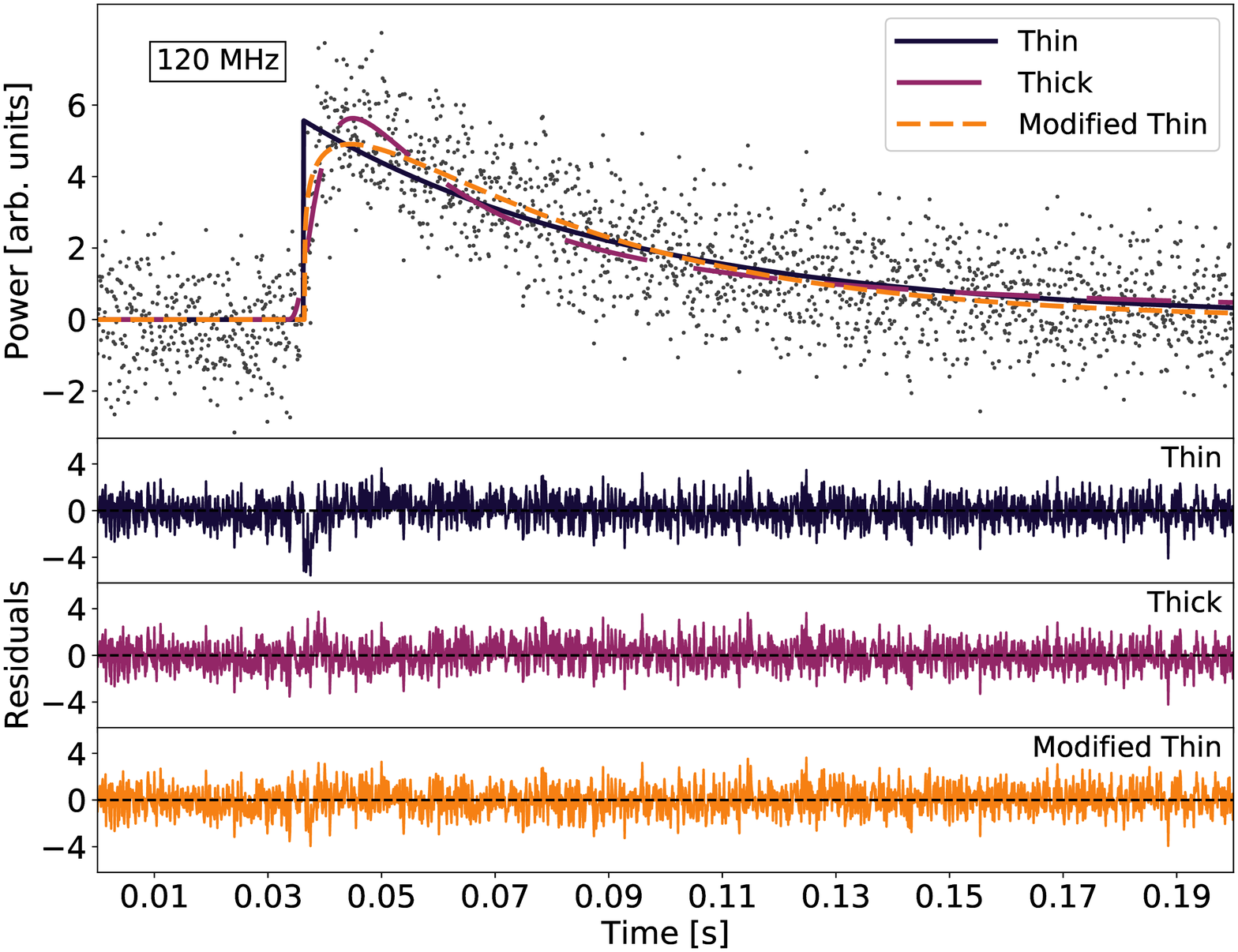}\includegraphics[width=0.5\textwidth]{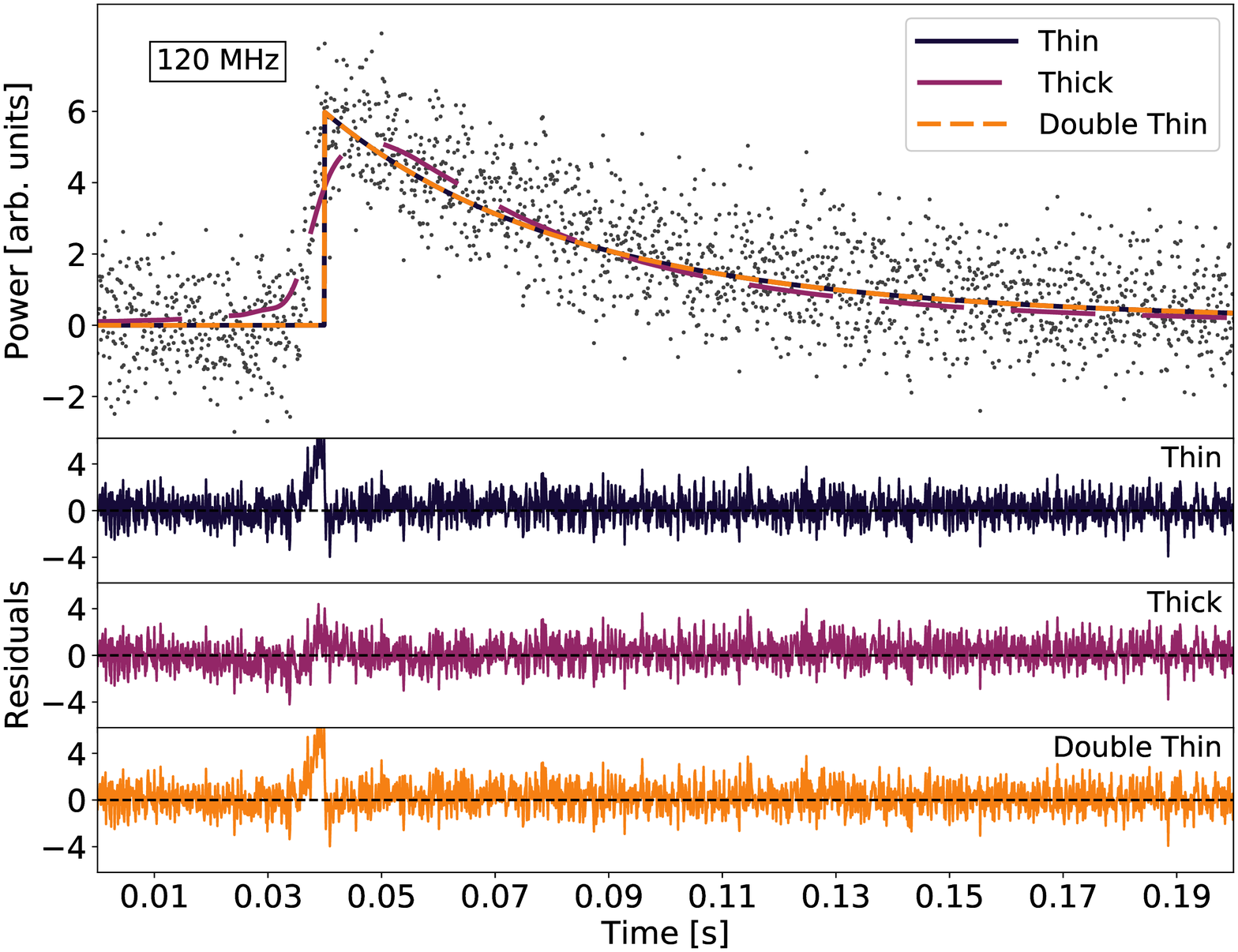} 
\par\end{centering}
\begin{centering}
\includegraphics[width=0.5\textwidth]{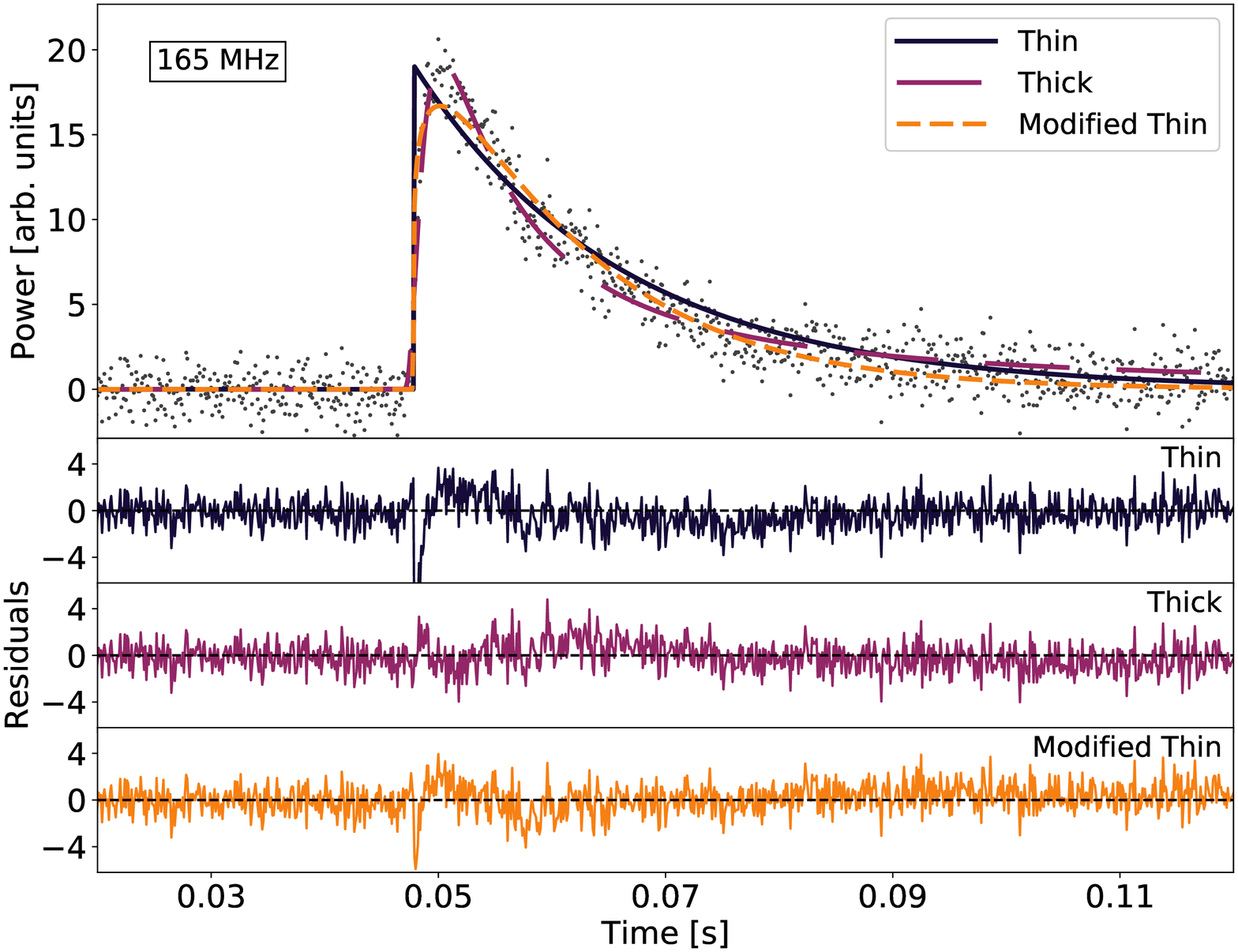}\includegraphics[width=0.5\textwidth]{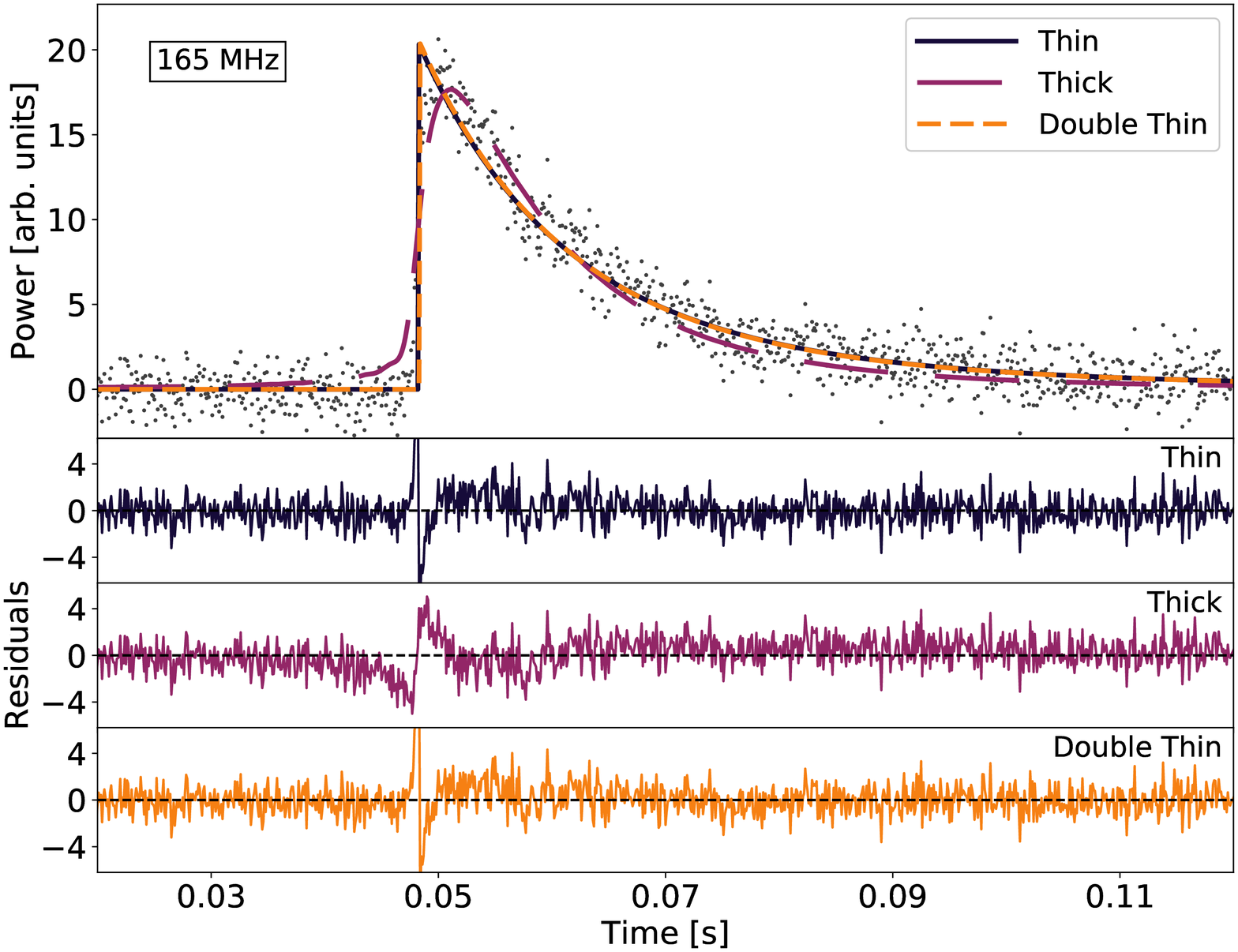} 
\par\end{centering}
\begin{centering}
\includegraphics[width=0.5\textwidth]{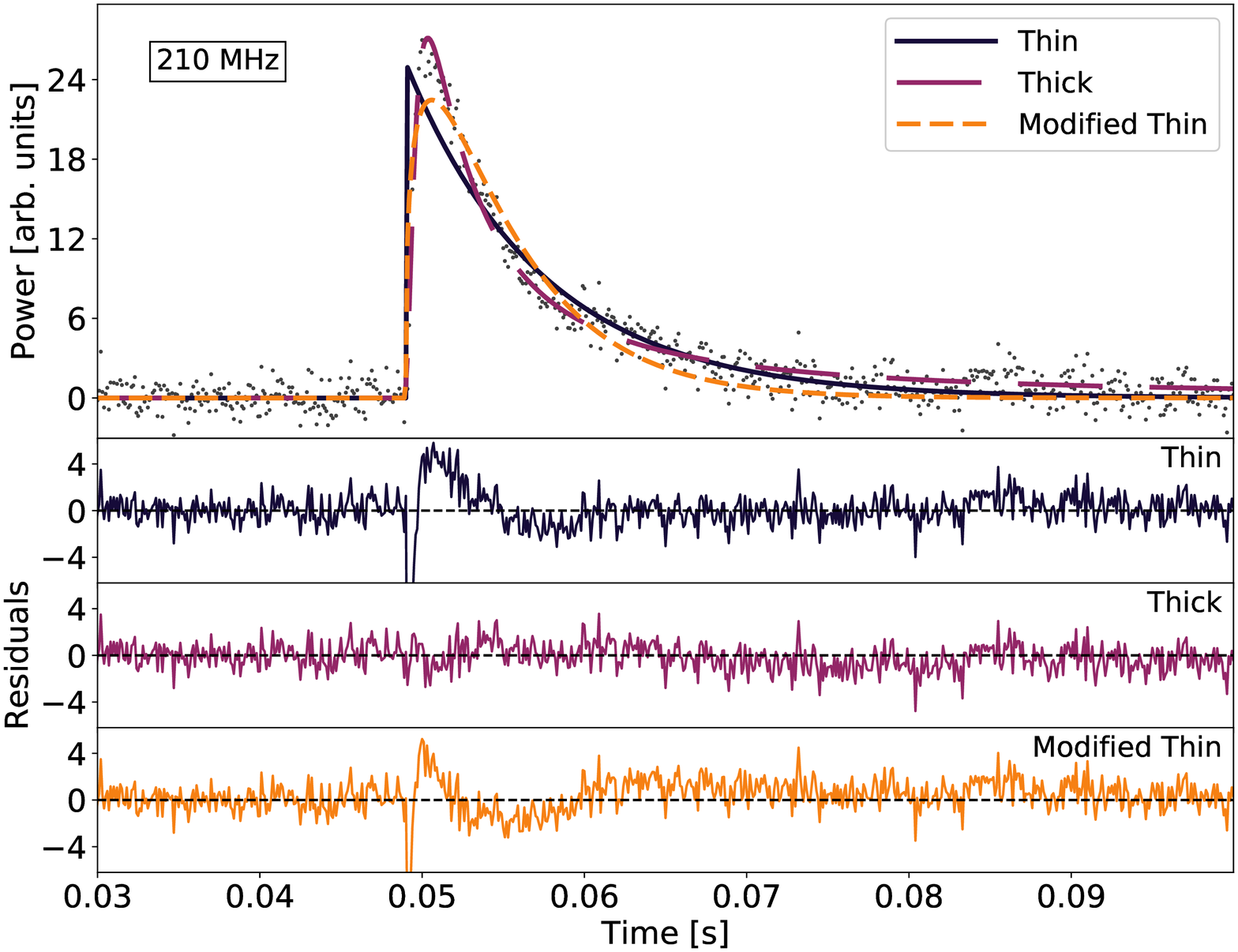}\includegraphics[width=0.5\textwidth]{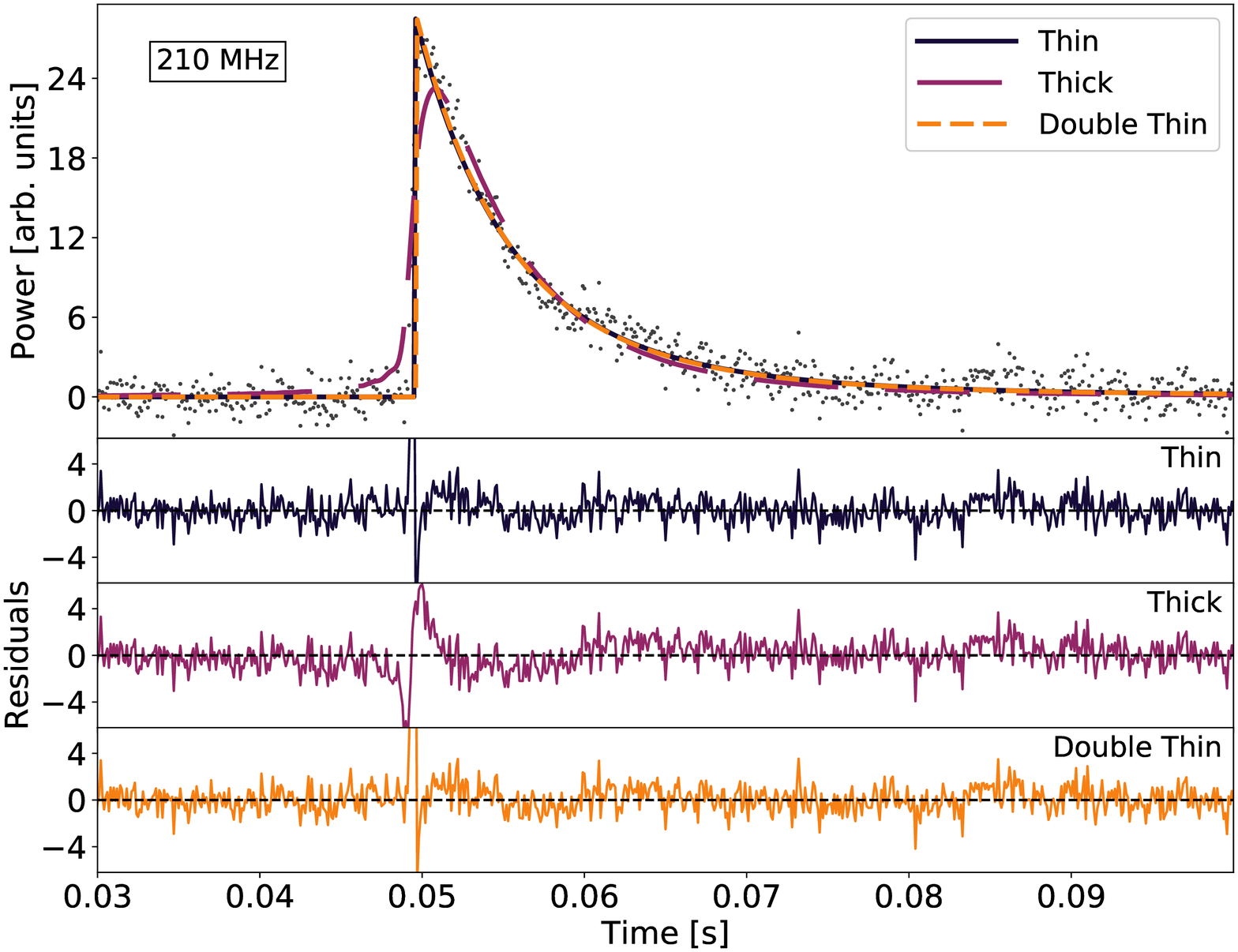} 
\par\end{centering}
\caption{\label{fig:Crab-Scattered-profiles-of}Scattered profiles of one particular
giant pulse of the Crab at 120~MHz (top row), 165~MHz (middle row),
210~MHz (bottom row) with different scattering models overplotted
and their respective residuals (color coded according to the model
below each pulse). The left column shows fits using the canonical
models of thin, thick and modified thin models for the scattering
screen while the right column show fits for fully diffractive models
of Kolmogorov turbulence. None of the models explains the observed
profile sufficiently well across all frequencies. Note the different
time scales at the different frequencies.}
\end{figure*}

The complexity of the pulse profile of the Crab pulsar and the degree
of scattering at MWA frequencies render the analysis of average profiles
unsuitable for the estimation of its scattering delays. The scattering
tail extends well beyond the pulsar period at the lower frequencies.

The Crab pulsar is known for giant pulses exceeding the flux density
of its regular pulses by orders of magnitude. Therefore, we can use
a single giant pulse to study the effects of the ISM on the Crab pulsar's
emission. Intrinsically, giant pulses are of $\sim\mu$s time duration
\citep[e.g.][]{bhat2008,popov2007}, so can be treated as impulsive
signals given the native resolution of our VCS recordings ($100\,\mu$s).
Hence, the problem of an unknown intrinsic pulse shape does not apply
to giant pulses and a normal least-squares fitting routine can be
applied when fitting scattering models. We tried both PBF forms in
Equations \ref{eq:thinscreen} and \ref{eq:thickfinite} and, additionally,
also fit a modified exponential function defined as

\begin{equation}
g_{2}(t)=t^{\gamma}\exp\left[-\frac{t}{\tau}\right]u(t),\label{eq:assymexp}
\end{equation}
where $\gamma$ is a free parameter in the range $0\mathrm{-}1$ describing
the rise time of the pulse; $u(t)$ is as defined for Equation \ref{eq:thinscreen}.
Both \citet{2012A&AKaruppusamy} and \citet{2013ApJEllingson} have
made use of this PBF and found that this functional form fits their
data better at low frequencies. We restrict our analysis to the lower
three bands where the signal-to-noise ratio (SNR) is highest (Fig.
\ref{fig:vela+crab-stacked-freq}). The data and associated fits are
shown in in the left column of Figure \ref{fig:Crab-Scattered-profiles-of}
and the measured scattering delays along with the reduced $\chi_{\mathrm{red}}^{2}$
values are summarized in Table \ref{tab:crab-fitted-delays}.

Figure \ref{fig:Crab-Scattered-profiles-of} (right column) also contains
the results of numerical fits where we employed PBFs that are based
on a fully diffractive Kolmogorov model of turbulence \citep[e.g.][]{lambert1999}.
This work is motivated by the fact that none of the above three models
(Eqs. \ref{eq:thinscreen}, \ref{eq:thickfinite}, and \ref{eq:assymexp}),
which are based on the assumption of a square-law structure function
of the ISM, fits the data well across all three bands. Moreover, all
three models assume that the scattering occurs in only one location
along the path. For the Crab pulsar it has been suggested there may
be two distinct contributing scattering locations, namely material
within the Crab nebula itself and the ISM between Earth and the pulsar
\citep{vandenberg76-B}. Therefore, we are motivated to explore a
set of models in which the scattering occurs at multiple locations
along the ray path. In particular, we investigate the case in which
the scattering occurs on multiple thin screens. It may be shown that
the multiple thin screen geometry gives rise to a scattering kernel
whose shape is the convolution of the scattering kernels of the two
screens individually. 

The scattering delays that we measure for each of the six models indicate
a frequency scaling $-4.0<\alpha<-3.4$ (Table \ref{tab:crab-fitted-delays}).
This is slightly steeper than but consistent with the the results
of, e.g., \citet{meyers2017,eftekhari2016}, and in excellent agreement
with a global scattering index $\langle\alpha\rangle=-3.9\pm0.2$
determined by \citet{bhat2004}. The parameter $\gamma$ in Eq. \ref{eq:assymexp}
evaluates to $0.18\pm0.02$ at all frequencies. This is in good agreement
with the values found by \citet{2012A&AKaruppusamy} and lower but
still within three standard deviations of the results found by \citet{2013ApJEllingson}.

\section{Discussion\label{sec:Discussion}}

\subsection{PSR J0742$-$2822 }

\textbf{}
\begin{table}
\textbf{\caption{\label{tab:pulsar-residual-smearing-times}Residual dispersion smearing
time across $10\,$kHz at $120\,$MHz and $230\,$MHz}
}

\textbf{}%
\begin{tabular}{cccc}
 & DM & \multicolumn{2}{c}{$\tau_{\mathrm{disp}}${[}ms{]}}\tabularnewline
Pulsar & {[}pc$\,$cm$^{-3}${]} & $120\,$MHz & $230\,$MHz\tabularnewline
\hline 
J0534$+$2200  & $56.7$ & $2.72$ & $0.39$\tabularnewline
J0742$-$2822  & $73.7$ & $3.54$ & $0.50$\tabularnewline
J0835$-$4510 & $67.9$ & $3.26$ & $0.46$\tabularnewline
\hline 
\end{tabular}
\end{table}

The work of \citet{johnston98} suggested that the interstellar scattering
along this line of sight is dominated by a thin screen located within
the Gum Nebula. Our measurements are well in line with this notion
as the FOM (Table \ref{tab:0742-scattering-times}) favor the thin
screen model over the thick screen one. For all eight frequency bands
considered in the analysis, both $f_{r}$ and $\chi_{\mathrm{nCC}}^{2}$
are lower for the thin screen model than the thick screen one, indicating
a better fit to the data. The thin screen model yields a scaling index
$\alpha=-2.5\pm0.6$ that is in agreement with previous work \citep[e.g.][]{lewandowski2015}.
However, the measured scattering delays $\tau$ and the frequency
scaling $\alpha=-2.9\pm0.7$ as obtained from assuming a thick screen
model agree with the thin screen estimates within one standard deviation. 

One can also compare the recovered deconvolved pulse shapes between
models with those obtained at higher frequencies. We show our profiles
in Fig. \ref{fig:reconstructed pulses} (left) which hint at a double-peaked
pulse regardless of the assumed underlying model for scattering. Similar
double peaked profiles with matching component separations are evident
at higher frequencies for this pulsar \citep[e.g.][and references therein]{kramer1994,zhao2017}.
However, as the signal-to-noise ratio (SNR) in the individual subbands
is low, these reconstructed pulse shapes should be considered as first
order estimates only. The measured scattered profiles (Fig. \ref{fig:0742-stacked-freq},
left), on the other hand, are very reliable as they were constructed
from about $1800$ individual pulses. 

Furthermore, we can consider the scintillation properties at higher
frequencies implied by a certain frequency scaling. For diffractive
scintillation, the decorrelation bandwidth $\Delta\nu_{\mathrm{d}}$
is related to the scatter broadening time $\tau$ via the relation
$\Delta\nu_{\mathrm{d}}=C/(2\pi\tau)$, where $C$ is of order unity
\citep[$C$ ranges from 0.65 to 1.2 depending on assumed geometry and underlying density spectrum]{lambert1999}.
For PSR J0742$-$2822 an $\alpha=-2.5$ (or $\alpha=-2.9$ for the
thick screen) would predict a decorrelation bandwidth $\Delta\nu_{\mathrm{d}}\approx0.1\,$MHz
($\Delta\nu_{\mathrm{d}}\approx0.2\,$MHz, thick screen) at $4.8\,$GHz
that is only about an order of magnitude lower than what has been
measured by \citet{johnston98} at that frequency ($\Delta\nu_{\mathrm{d}}=8.83\,$MHz).
Considering the time variability of scintillation and the uncertainties
of our measurements our results are well in line with these earlier
findings. 

\subsection{The Vela Pulsar}

\begin{figure*}
\includegraphics[width=1\textwidth]{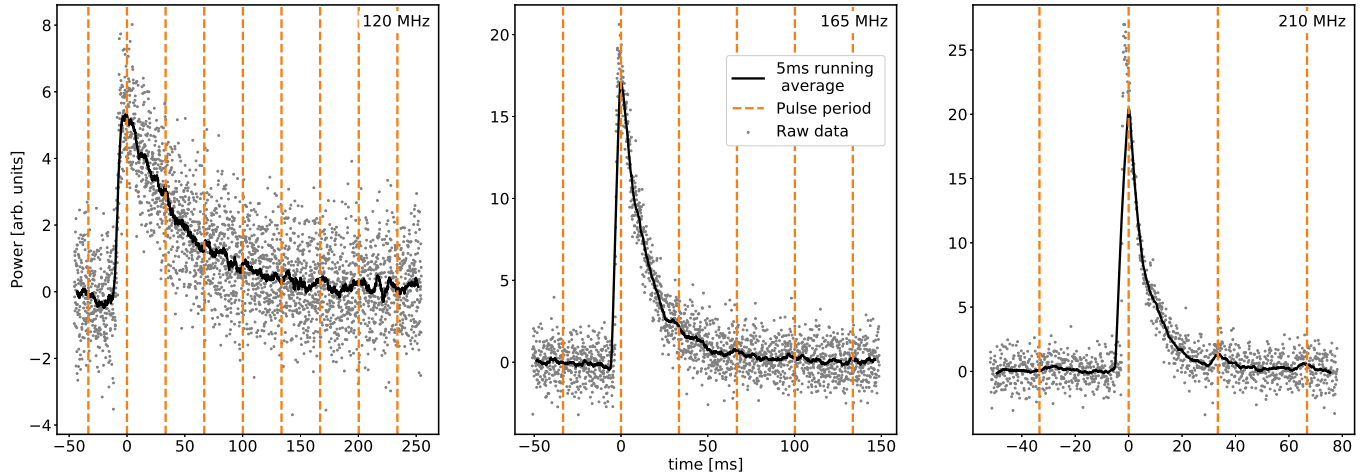}

\caption{\label{fig:crab-pulse-runningAv}Raw data (gray dots) and a five milliseconds
running average thereof (black line) of the Crab giant pulse fitted
in Figure \ref{fig:Crab-Scattered-profiles-of}. The vertical yellow
dashed lines indicate time steps of one pulse period of the Crab ($\approx33.3\,$ms)
relative to the peak of the averaged profile. Note how the power in
the scattering tail is increased by pulses succeeding the giant pulses,
ultimately influencing the results of the fitting procedure. Also
note the different time ranges at the different central frequencies
of $120\,$MHz (left), $165\,$MHz (center), and $210\,$MHz (right).}
\end{figure*}

Similar to PSR J0742$-$2822, it has been speculated that the Gum
Nebula is the dominant source of scattering along the line of sight
to the source \citep{backer74}, favoring the model of a thin scattering
screen over a thick one. The measurements we present in Table \ref{tab:vela-scattering-times},
however, are inconclusive as neither the frequency scaling nor the
FOM enable us to critically distinguish between the two interstellar
scattering models used in the analysis. To a large degree this is
caused by the limited amount of data points (the SNR was high enough
in only the highest two out of five subbands) and the fact that the
we are limited in both time and frequency resolution ($100\,\mu$s
and $10\,$kHz, respectively) leading to (i) the inability to recover
small scale pulse structure \citep[e.g.][for the Vela pulsar]{johnston2001};
and (ii) residual dispersive smearing (Table \ref{tab:pulsar-residual-smearing-times}).
The overall effect is that the recovered, deconvolved pulse shapes
are a low-resolution estimate of the intrinsic pulse shapes, hindering
discrimination between scattering models (Fig. \ref{fig:reconstructed pulses},
right). With higher time and frequency resolution one could compare
the recovered pulse shapes with those observed at higher frequencies,
providing a further aspect against which to compare the goodness of
fit.

Within their respective uncertainties, both the thick and the thin
screen model yield a scaling index $\alpha$ that is in agreement
with previous work \citep[e.g.][]{lewandowski2015,johnston98}, which
fits into the standard picture of Kolmogorov turbulence. However,
one can again consider the implications of the scaling indices for
the decorrelation bandwidth at higher frequencies. For the Vela a
frequency scaling of $\alpha=-5.0$ would indicate a decorrelation
bandwidth $\Delta\nu_{\mathrm{d}}\approx1\,$MHz at an observing frequency
of $2.3\,$GHz, while a scaling $\alpha=-4.0$ would yield $\Delta\nu_{\mathrm{d}}\approx70\,$kHz.
The latter is in good agreement with $\Delta\nu_{\mathrm{d}}=66\,$kHz
measured by \citet{gwinn2000}. Thus, there is a preference for a
thin screen scattering geometry along the line of sight to the Vela
pulsar.

\subsection{The Crab\label{subsec:discussion-The-Crab}}

None of the models that we fitted to the brightest giant pulse in
our dataset reproduce the observed pulse shapes well across all three
frequency bands (Fig. \ref{fig:Crab-Scattered-profiles-of}). They
generally fail to capture the rise times correctly, with the model
for a thick scattering screen describing the leading edge best, i.e.
the residuals show nearly no deviation from flatness in this particular
time range of the pulse. However, this model, like all others, does
not reproduce the shape of the scattering tail in a way that the residuals
show no systematic offset from zero. At a central frequency of $165\,$MHz
the model for a thick scattering screen decays too steeply, i.e. it
does not account for all the power still present right after the peak
of the pulse, while at $210\,$MHz this model predicts more power
at later times in the tail than is actually present. The latter might
indicate the presence of an inner scale of turbulence as observed
by, e.g., \citet{rickett2009}. 

Similar to the analytic models, none of the numerical models we employed
appropriately characterize the immediate rise time of the pulse. Compared
to the analytic models, the residuals of the single thin and double
thin models are, however, somewhat flatter both right after the peak
of the pulse and at later times in the scattering tail. We interpret
this as an indication that compared to the analytical models, the
numerical models for thin screens better account for both the small
scale scattering structures as well as the larger scale structures.

The fitting results of both the analytic and the numerical approach
are strongly influenced by the fact that the scattering tail extends
beyond a single pulse period. This becomes evident in Figure \ref{fig:crab-pulse-runningAv}
where we plot the raw data in all three subbands alongside with a
five millisecond running average. In all three subbands one can identify
an increase in power in the scattering tail at intervals equal to
the pulse period. We interpret these peaks as pulses succeeding the
giant pulse, ultimately leading to a measurement of the scatter broadening
time that is larger than that induced by the ISM.

Similarly, the influence of subsequent pulses makes it virtually impossible
to distinguish between a simple exponentially decaying scattering
tail and one that decays slower as is expected for Kolmogorov turbulence
\citep{lambert1999}.

\section{Summary \& Conclusions}

\begin{figure*}
\begin{centering}
\includegraphics[width=1\textwidth]{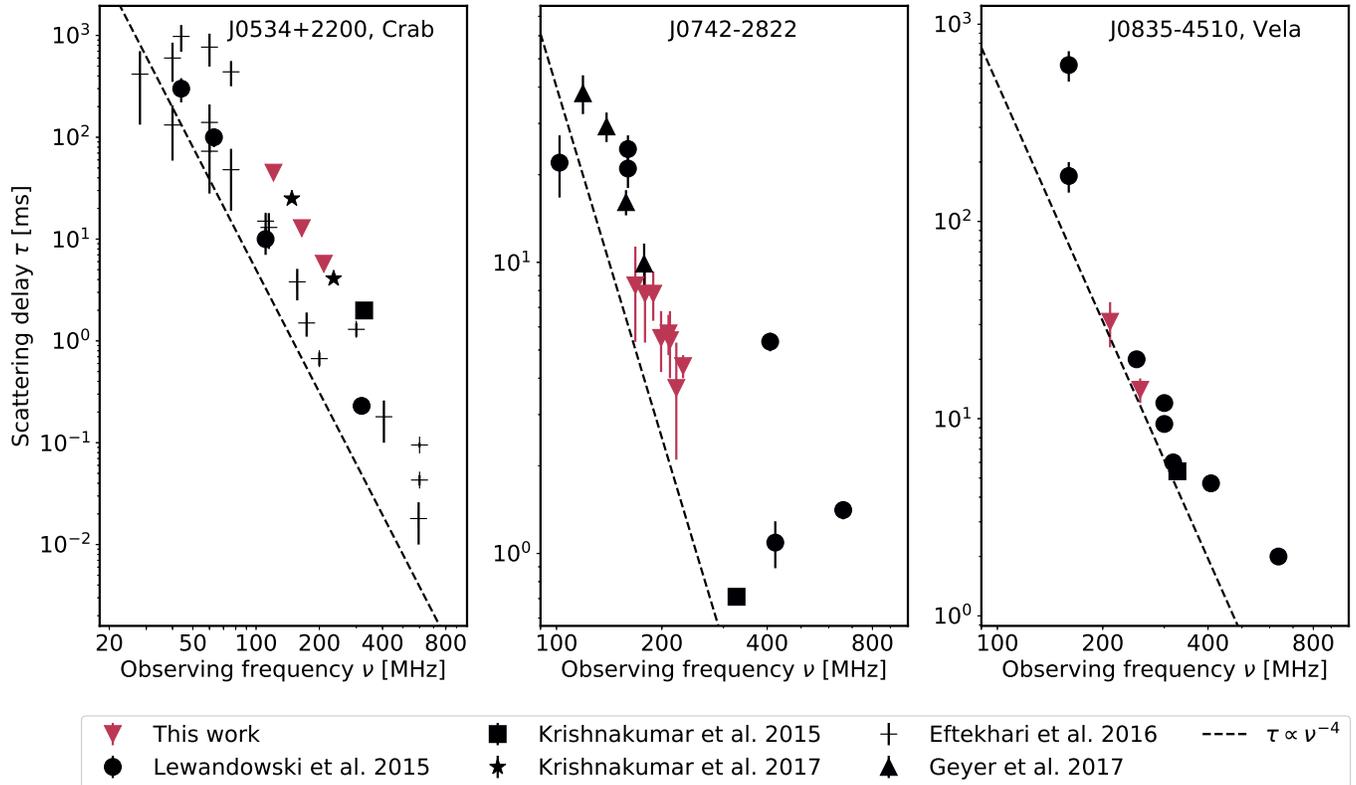}
\par\end{centering}
\caption{\label{fig:literature-vs-our-delays}Scattering delays below an observing
frequency $\nu<1000\,$MHz as found in the literature (black circles,
squares, stars, and triangles) and our measured values (red triangles)
for J0534$+$2200 (left, from thick screen model), J0742$-$2822 (middle,
from thin screen model) and J0835$-$4510 (right, from thin screen
model). All values from the literature are based on the assumption
of a single thin scattering screen (Eq \ref{eq:thinscreen}), with
the exception of the data from \citet{eftekhari2016} who used the
modified thin screen model (Eq. \ref{eq:assymexp}). The dashed line
indicates the canonical relation \textbf{$\tau\propto\nu^{-4}$.}}
\end{figure*}

We observed the three pulsars J0534$+$2200 (the Crab pulsar), J0742$-$2822,
and J0835$-$4510 (the Vela pulsar) across the entire frequency range
available with the MWA ($80-300\,$MHz). We employed both analytical
and numerical models for interstellar scattering to measure the pulse
scatter broadening times induced by the ISM between frequencies of
$148\,$MHz and $256\,$MHz. Based on these measurements we were able
to estimate the frequency scaling of interstellar scattering implied
by the different models which generally agree within their uncertainties
despite the fact that the measured $\tau$ differ significantly between
models. We find that both a thin and a thick screen model are consistent
with our data, albeit a preference for a thin screen geometry for
both J0742$-$2822 and J0835$-$4510. The thin screen models imply
frequency scalings $\alpha=-2.5\pm0.6$ and $\alpha=-4.0\pm1.5$ for
J0742$-$2822 and the Vela pulsar, respectively. These numbers are
in agreement with previous publications \citep[e.g.][]{lewandowski2015}.
In the case of the Crab pulsar our data indicate that none of our
employed models represents the data accurately across all observed
frequencies, with the model for a thick scattering screen fitting
the rise times of the analyzed giant pulse best. The frequency scaling
implied by this model is $\alpha=-3.8\pm0.2$. Overall, the spectral
scaling indices we measure are all shallower than what is expected
for pure Kolmogorov turbulence ($\alpha=-4.4$). The existence of
an inner scale of turbulence could at least partially explain this
discrepancy \citep[e.g.][]{rickett2009,bhat2004}.

The scattering delays that we measured in these short, multi-band,
single epoch observations with the MWA fit very well into what has
been measured previously (Figure \ref{fig:literature-vs-our-delays}).
In these previous works, the large scatter of the measured delays
at similar frequencies is most likely caused by combining data from
(i) separate observations spanning long time ranges, (ii) multiple
dissimilar telescopes, and (iii) varying techniques to measure the
scattering delay. Effectively, this will not only affect the measured
delays but also the implied frequency scaling index of pulsar scattering,
$\alpha$. In this work, we accomplished these measurements from relatively
short, single-epoch observations with the MWA alleviating all of the
above difficulties. This is a testimony to the sensitivity and flexibility
of the MWA and it underscores the importance of pulsar studies at
low radio frequencies.

\acknowledgements{We thank the two anonymous referees for insightful reviews and for
several comments that helped to improve the content and presentation
in this paper. FK acknowledges funding through the Australian Research
Council grant DP140104114 and from the Swedish Research Council. NDRB
acknowledges the support from a Curtin Research Fellowship (CRF12228).
The authors acknowledge the contribution of an Australian Government
Research Training Program Scholarship in supporting this research.
This scientific work makes use of the Murchison Radio-astronomy Observatory,
operated by CSIRO. We acknowledge the Wajarri Yamatji people as the
traditional owners of the Observatory site. Support for the operation
of the MWA is provided by the Australian Government (NCRIS), under
a contract to Curtin University administered by Astronomy Australia
Limited. We acknowledge the Pawsey Supercomputing Centre which is
supported by the Western Australian and Australian Governments. Parts
of this research were conducted by the Australian Research Council
Centre of Excellence for All-sky Astrophysics (CAASTRO), through project
number CE110001020.}

\facilities{MWA}

\software{PSRCHIVE, DSPSR, PYTHON, MATLAB}

\appendix

\section{Figures of Merit\label{app:Figures-of-Merit}}

\begin{figure*}
\includegraphics[width=1\textwidth]{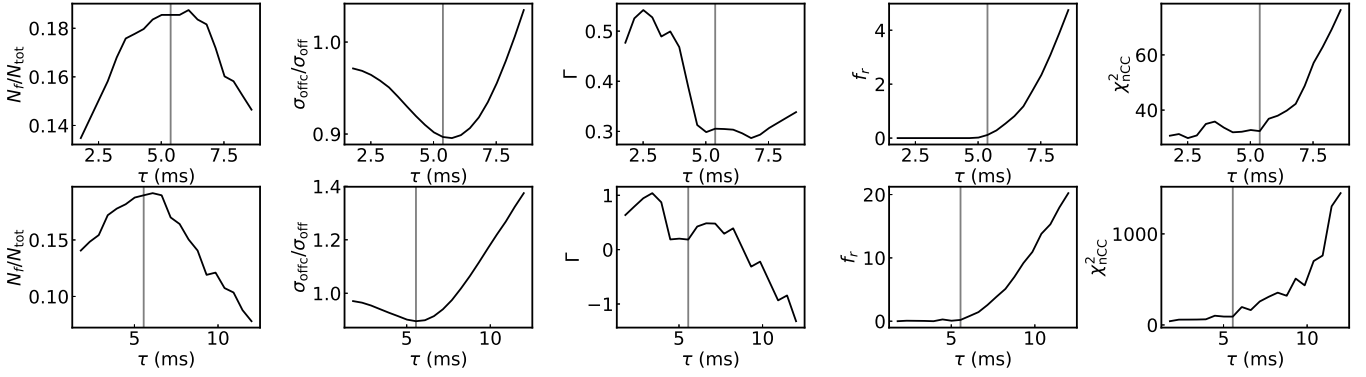}

\caption{\label{fig:Examples-of-the-fom}Examples of the collection of figures
of merit employed to determine the best fit model. Shown are the FOMs
as computed for PSR J0742$-$2822 at an observing frequency of $211.2\,$MHz.
Top row: FOMs for thin screen model (Eq. \ref{eq:thinscreen}); bottom
row: FOMs for thick screen model (Eq. \ref{eq:thickfinite}). The
gray horizontal lines indicate the value of $\tau$ that was taken
to be the best fit value.}
\end{figure*}

Here we summarize the definitions of the FoM as defined in \citep{bhat03}.
An example of how we chose the best fit model is shown in Fig. \ref{fig:Examples-of-the-fom}.

The parameter $f_{\text{r}}$ is defined as
\[
f_{r}=\frac{1}{N\sigma_{\mathrm{off}}^{2}}\sum_{i=1}^{N}[\Delta y(t_{i})]^{2}U_{\Delta y},
\]
where $U_{\Delta y}$ is the unit step function defined such that
\[
U_{\Delta y}=\begin{cases}
1 & \mathrm{if}\;(\Delta y(t_{i})+1.5\sigma_{\mathrm{off}})<0,\\
0 & \mathrm{otherwise,}
\end{cases}
\]
and $N$ is the total number of time bins $t_{i}$ during which the
pulse is 'on'. $\sigma_{\mathrm{off}}$ denotes the off-pulse rms
noise while the parameter $\Delta y(t_{i})$ denotes the value of
time bin $t_{i}$ in the CLEAN residual (i.e. the residual profile
after subtraction of all CLEAN components). Effectively, $f_{r}$
is a measure for positivity, i.e. how much oversubtraction occurs
given the CLEANed pulse of a certain model.

$N_{\text{r}}=N_{\mathrm{f}}/N_{\mathrm{tot}}$ denotes the relative
number of points $N_{\mathrm{f}}$ that satisfy $|y_{i}-\langle y_{\text{off}}\rangle|\leq3\sigma_{\text{off}}$,
where $\langle y_{\text{off}}\rangle$ denotes the off-pulse mean
signal and $N_{\mathrm{tot}}$ are the total number of points in the
profile. 

$\sigma_{\mathrm{r}}=\sigma_{\mathrm{offc}}/\sigma_{\mathrm{off}}$
is the ratio between the rms of the residual after deconvolution and
the off-pulse rms in the scattered profile. It is a relative measure
for how noise-like the residual of the deconvolved pulse is. 

The skewness $\Gamma$ is a measure for the symmetry of the CLEANed
profile and is computed as 

\textbf{
\begin{align*}
\Gamma= & \frac{\langle t^{3}\rangle}{\langle t^{2}\rangle^{3/2}},\,\mathrm{with}\\
\langle t^{n}\rangle= & \frac{\sum_{i=1}^{n_{c}}(t_{i}-\bar{t})^{n}C_{i}}{\sum_{i=1}^{n_{c}}C_{i}}\,,\\
\bar{t}= & \frac{\sum_{i=1}^{n_{c}}t_{i}C_{i}}{\sum_{i=1}^{n_{c}}C_{i}}\,.
\end{align*}
}

Here, the summation parameter $i=1,...,n_{c}$ runs through all CLEAN
components of amplitudes $C_{i}$ and associated times $t_{i}$.

\bibliographystyle{aasjournal}
\bibliography{biblio}

\end{document}